\documentclass[10pt,twocolumn,letterpaper]{article}

\usepackage{cvpr}              
\usepackage{graphicx}
\usepackage{amsmath}
\usepackage{amssymb}
\usepackage{booktabs}
\usepackage{multirow}
\usepackage{times}
\usepackage{algorithm}
\usepackage{algorithmic}
\usepackage{cite}
\usepackage{fancyvrb}

\usepackage[table]{xcolor}
\usepackage{xcolor} 
\definecolor{Gray}{gray}{0.9}

\usepackage[pagebackref,breaklinks,colorlinks]{hyperref}

\usepackage[capitalize]{cleveref}
\crefname{section}{Sec.}{Secs.}
\Crefname{section}{Section}{Sections}
\Crefname{table}{Table}{Tables}
\crefname{table}{Tab.}{Tabs.}

\def\cvprPaperID{10132} 
\def\confName{CVPR}
\def\confYear{2023}

{\begin{list}               
    {$\bullet$ \hfill}{
        \setlength{\leftmargin}{\parindent}
        \setlength{\parsep}{0.04\baselineskip}
        \setlength{\itemsep}{0.5\parsep}
        \setlength{\labelwidth}{\leftmargin}
        \setlength{\labelsep}{0em}}
    }
{\end{list}}

\providecommand{\cref}[1]{Chapter~\ref{#1}}

\providecommand{\fref}[1]{Figure~\ref{#1}}

\providecommand{\R}{\ensuremath{\mathbb{R}}}

\providecommand{\calN}{\mathcal{N}}

\providecommand{\calP}{\mathcal{P}}

\providecommand{\vx}{\mathbf{x}}

\begin{document}

\title{HDR Imaging with Spatially Varying Signal-to-Noise Ratios}

\author{Yiheng Chi \hspace{10ex} Xingguang Zhang \hspace{6ex} Stanley H. Chan \\
School of Electrical and Computer Engineering\\
Purdue University\\
{\tt\small chi14@purdue.edu}  \hspace{5ex} {\tt\small zhan3275@purdue.edu} \hspace{2.5ex} {\tt\small stanchan@purdue.edu}
}

\maketitle

\begin{abstract}
While today's high dynamic range (HDR) image fusion algorithms are capable of blending multiple exposures, the acquisition is often controlled so that the dynamic range within one exposure is narrow. For HDR imaging in photon-limited situations, the dynamic range can be enormous and the noise within one exposure is spatially varying. Existing image denoising algorithms and HDR fusion algorithms both fail to handle this situation, leading to severe limitations in low-light HDR imaging.

This paper presents two contributions. Firstly, we identify the source of the problem. We find that the issue is associated with the co-existence of (1) spatially varying signal-to-noise ratio, especially the excessive noise due to very dark regions, and (2) a wide luminance range within each exposure. We show that while the issue can be handled by a bank of denoisers, the complexity is high. Secondly, we propose a new method called the spatially varying high dynamic range (SV-HDR) fusion network to simultaneously denoise and fuse images. We introduce a new exposure-shared block within our custom-designed multi-scale transformer framework. In a variety of testing conditions, the performance of the proposed SV-HDR is better than the existing methods.
\end{abstract}

\section{Introduction}
Today's high dynamic range (HDR) image fusion algorithms have demonstrated remarkable performances in blending images across a wide range of luminance levels. Many algorithms are able to handle an interior room with a sunlit view, of which the overall dynamic range is in the order of 100000:1 or more. However, most of these algorithms are designed for well-illuminated scenes. Even in the shortest exposure frame, the noise is maintained at a modest level so that the algorithm can focus on the blending task. The question we ask in this paper is: What if we push the shortest exposure to a photon-starving condition?

\begin{figure}[t]
\begin{tabular}{cccc}
\multicolumn{4}{c}{\includegraphics[width=0.95\linewidth]{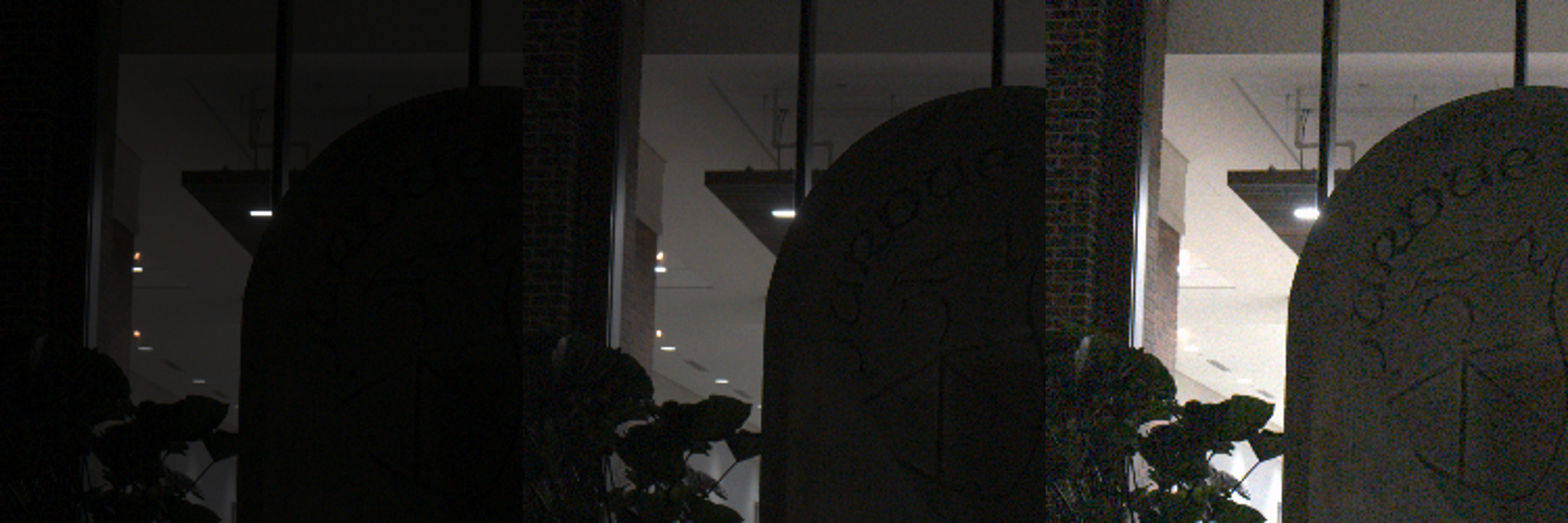}}\\
\includegraphics[width=0.23\linewidth]{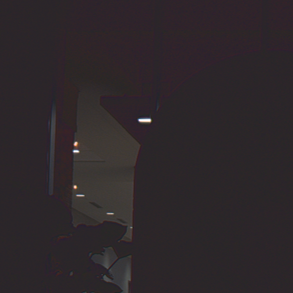} &
\hspace{-2ex}\includegraphics[width=0.23\linewidth]{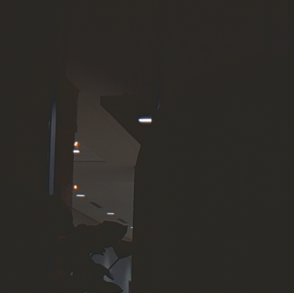} &
\hspace{-2ex}\includegraphics[width=0.23\linewidth]{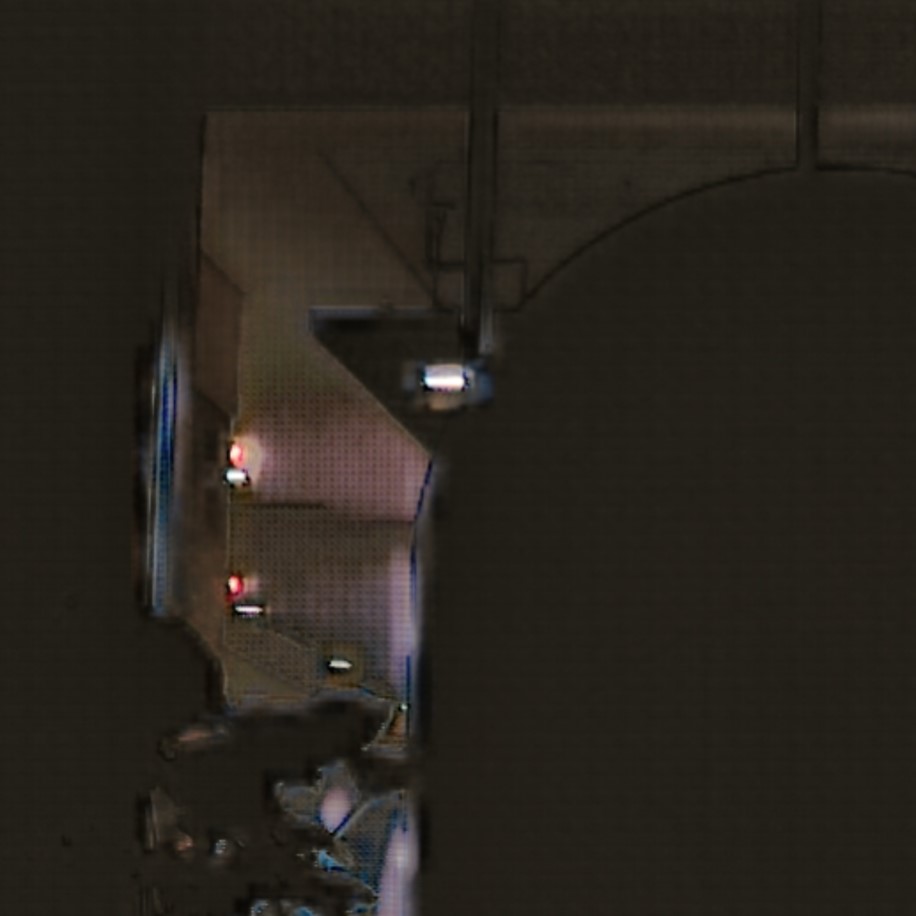} &
\hspace{-2ex}\includegraphics[width=0.23\linewidth]{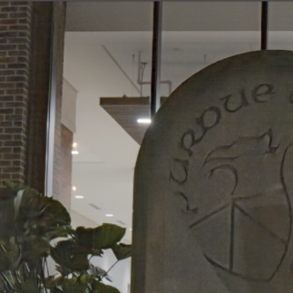}\\
\footnotesize{Kalantari \cite{kalantari2017deep}} & \hspace{-2ex} \footnotesize{Wu \cite{wu2018deep}} & \hspace{-2ex} \footnotesize{NHDRRNet \cite{yan2020deep}} & \hspace{-2ex} \footnotesize{Proposed} 
\end{tabular}
\caption{[Top] Real captures using a Sony ILCE-7M2 camera. Three low-dynamic range (LDR) images are captured. [Bottom] Image denoising and HDR fusion results.}
\label{fig: main figure 1}
\end{figure}

Such an extreme HDR problem arises in many low-light scenarios. \fref{fig: main figure 1} is a real image captured by a Sony ILCE-7M2 camera. The imaging condition is a night-time scenario in front of a building. The challenge of the problem is the co-existence of heavy noise in the darkest spots of the image and the high dynamic range. We refer to this as the spatially varying signal-to-noise ratio (SNR) problem where brighter pixels have higher SNR and darker pixels have lower SNR. 

The goal of this paper is to articulate the spatially varying SNR problem. We emphasize the difficulty of the problem by referring to the performance of three state-of-the-art HDR fusion algorithms, namely, Kalantari and Ramamoorthi \cite{kalantari2017deep}, Wu et al. \cite{wu2018deep}, and NHDRRNet \cite{yan2020deep}. As we can see from \fref{fig: main figure 1}, these methods produce disappointing results, mostly failing in denoising the dark regions.

\begin{figure}[t]
\centering
\includegraphics[width=0.9\linewidth]{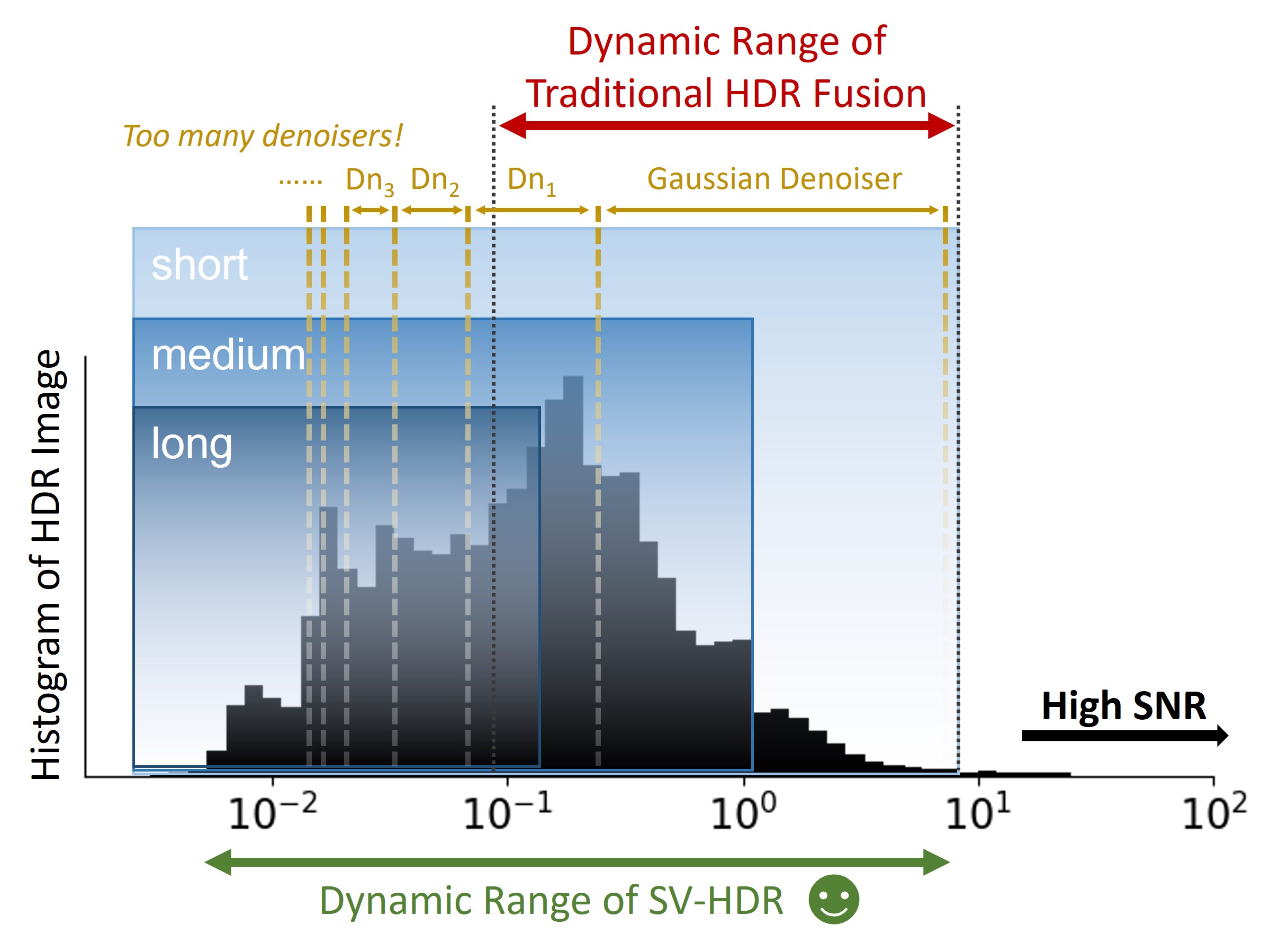}
\caption{Traditional HDR algorithm can handle high SNR cases. Individual denoisers Dn1, Dn2, Dn3 have narrow operating regimes. SV-HDR offers a wide dynamic range coverage with denoising capability.}
\label{fig: histogram}
\end{figure}

The position of the paper can be visualized in \fref{fig: histogram}. While existing HDR fusion methods can handle the blending task, the individual exposures are sufficiently high so that the amount of noise is limited. Single image denoisers today seldom handle the dynamic range problem. They are mostly focusing on a tonemapped image normalized to $[0,1]$. Therefore, when facing a wide dynamic range scene, we need multiple denoisers to denoise the images before blending them. The proposed method solves both the noise problem and the dynamic range problem at once.

The main contribution of this paper is a new HDR fusion and denoising network called the spatially varying high dynamic range network (SV-HDR). SV-HDR simultaneously denoises the image and blends three exposures into a single HDR image. Our network is a transformer-based approach with three customized designs: (1) A multi-exposure transformer block to extract features. These transformers are adaptive to the varying SNRs. (2) We introduce an exposure-share block to blend the features coming from the three exposures. (3) We incorporate a multi-scale blending strategy to capture the local and global variations.

\section{Problem Formulation and Related Work}
\label{sec: related work}
In this section, we present the problem formulation and discuss the related work. We also briefly discuss a failure analysis of existing methods.

\subsection{Realistic Image Sensor Model}
We start by discussing the image formation model. Unlike the standard denoising problems where we can assume a Gaussian/Poisson model, in this paper, we need a precise model to emulate the actual image formation process. 

A realistic image sensor model involves several parameters as determined by the following equation
\begin{align}
z(\vx)
&= \text{ADC}\left\{\text{Clip}\left\{\alpha  \times \calP(\tau \times \text{QE}  \times (\theta(\vx) + \mu_{\text{dark}})) \right\}\right\} \notag \\
&\qquad + \calN(0,\sigma_{\text{read}}^2).
\label{eq: main sensor model}
\end{align}
Here, the underlying scene flux is denoted by $\theta(\vx)$ where $\vx \in \R^2$ is the spatial coordinate. The observation is $z(\vx)$, which is pixel dependent. Other parameters such as the dark current and quantum efficiency, are listed in Table~\ref{table: sensor model}.

\begin{table}[h]
\caption{Image Sensor Model Parameters}
\label{table: sensor model}
\centering
\begin{tabular}{lll}
\hline
Symbol & Meaning & Typical Value\\
\hline
ADC & Analog-digital  & 14-bit \\
Clip& Full well limit & 5000 e- \\
$\tau$ & Exposure     & 50ms\\
$\mu_{\text{dark}}$ & dark current & 0.02 e-/s\\
QE  & Quantum efficiency & 50\%\\
$\alpha$ & Conversion gain & 1 at ISO 500 \\
$\sigma_{\text{read}}$ & Read noise & 10 ADU at ISO 500 \\
$\calP$ & Poisson distribution & \\
$\calN$ & Gaussian distribution & \\
\hline
\end{tabular}
\end{table}

The above equation is sufficient for the problem we are interested in. The model does not take into consideration of other ``high-order'' effects such as instability of the ADC thresholds, dead pixels (including non-responsive, overly sensitive, random telegraph signal, cross-talk, etc), pixel response non-uniformity, underflow offsets, and more. It also assumes a perfect color filter array, i.e., the electric and optical cross-talks are negligible.

\subsection{Related Work}
\textbf{Classical HDR Fusion Methods}. Debevec et al. \cite{debevec2008recovering} is one of the earliest papers that proposes combining multiple low dynamic range (LDR) images to construct an HDR image. Extending the idea to scenes that contain motion, people began to estimate the dynamic partial images and reject pixels \cite{khan2006ghost, grosch2006fast, jacobs2008automatic, pece2010bitmap, heo2010ghost, zhang2011gradient, lee2014ghost, oh2014robust, li2020fast}. Since pixels are dropped, these methods often suffer from loss of information. Another family of HDR fusion is registration-based, where pixels are first aligned with respect to a reference frame. The alignment can be done using optical flow \cite{zimmer2011freehand, bogoni2000extending, kang2003high}, energy optimization \cite{Sen2012Robust, hu2013hdr}, and rank minimization \cite{zheng2013hybrid}. These methods can handle small motions, but they still suffer from large foreground-object motions and they cannot manage noise. 

\textbf{Deep HDR Fusion Methods}. Deep learning approaches have been used on single-image HDR reconstructions such as \cite{eilertsen2017hdr, endo2017deep, santos2020single} and HDR video reconstructions such as \cite{kalantari2019deep}. There are also attempts at solving HDR fusion with large foreground motion. Kalantari and Ramamoorthi \cite{kalantari2017deep} propose a two-step fusion consisting of an optical flow estimation network and a fusion network. Wu et. al. \cite{wu2018deep} propose a single step method based on U-Net \cite{ronneberger2015u} and ResNet \cite{he2016deep}, where LDR frames are processed in different branches. Yan \cite{yan2019attention} brings the work forward by proposing a redesigned merging network and attention modules. Using attention to extract features is effective. The idea is adopted by various groups, e.g., \cite{yan2022lightweight, deng2020multi, zhu2022hdrfeat}. Apart from attention, \cite{yan2020deep} uses a non-local network to solve the HDR fusion problem, and \cite{yan2019multi} uses optical flow with a multi-scale dense network. 

\textbf{Low light HDR with New Sensors}. With the proliferation of Quanta Image Sensor (QIS) and Single Photon Avalanche Diodes (SPAD), people began to explore new sensors for HDR in low light. \cite{Gnanasambandam_TCI_HDR} and \cite{Chan_2022_TCI} showed the theoretical performance of QIS, whereas \cite{della2017high, instruments3030038, Gupta_Passive_2019, Gupta_QuantaBurst_2020} showed the corresponding result using SPAD. There is also a new dual-exposure sensor reported by \cite{ccougalan2020hdr} which demonstrates denoising and deblurring for HDR video.

\subsection{Why do Denoisers Fail?} \label{preliminary experiment}
As illustrated in \fref{fig: main figure 1}, existing HDR fusion algorithms fail to blend images captured in the nighttime. Our hypothesis is that it is the co-existence of the shot noise and the dynamic range that causes the difficulty. In this subsection, we confirm this speculation through two ablation studies.

\begin{figure}[h]
\begin{tabular}{ccc}
\hspace{0ex} \includegraphics[width=0.32\linewidth]{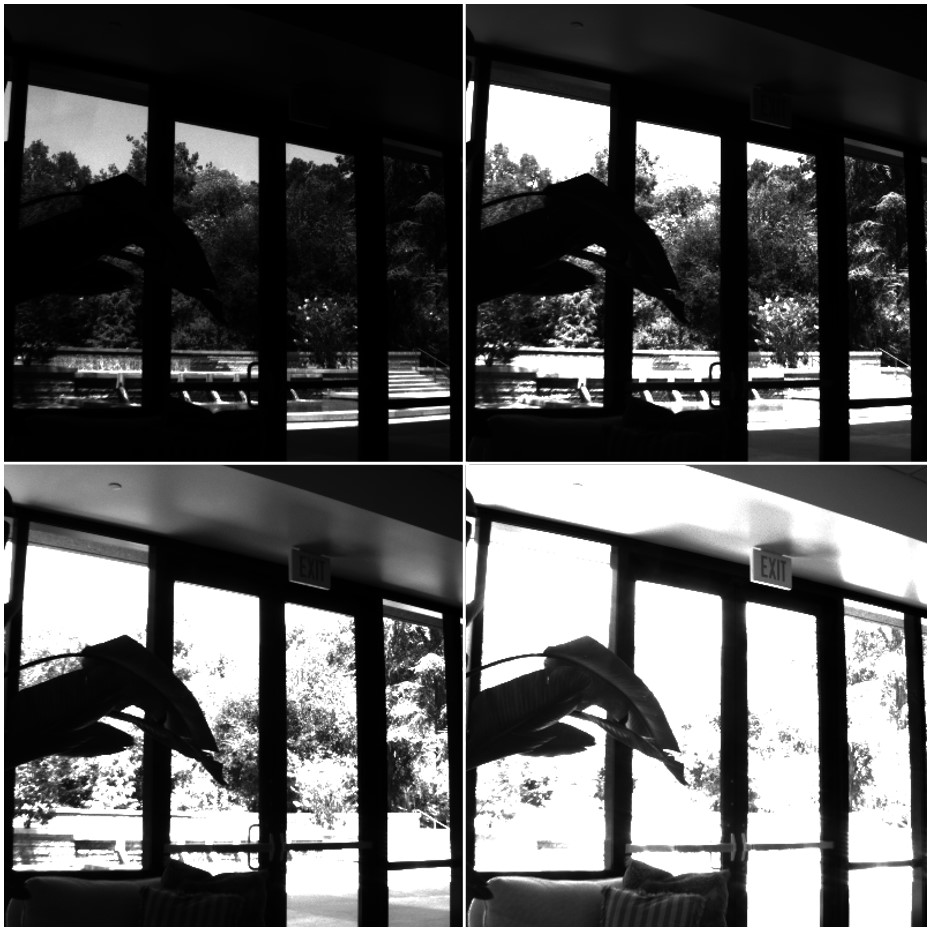}&
\hspace{-2.7ex} \includegraphics[width=0.32\linewidth]{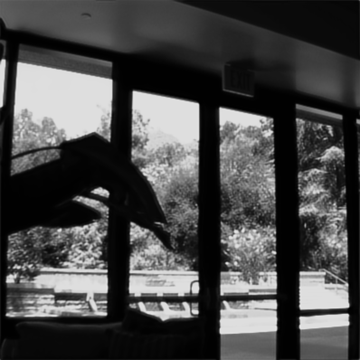}&
\hspace{-2.7ex} \includegraphics[width=0.32\linewidth]{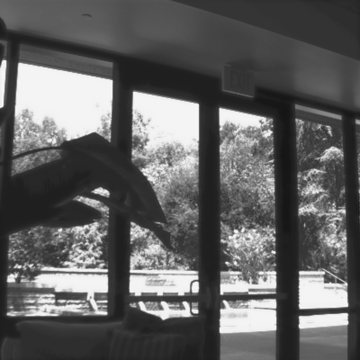}\\
\hspace{0ex} \includegraphics[width=0.32\linewidth]{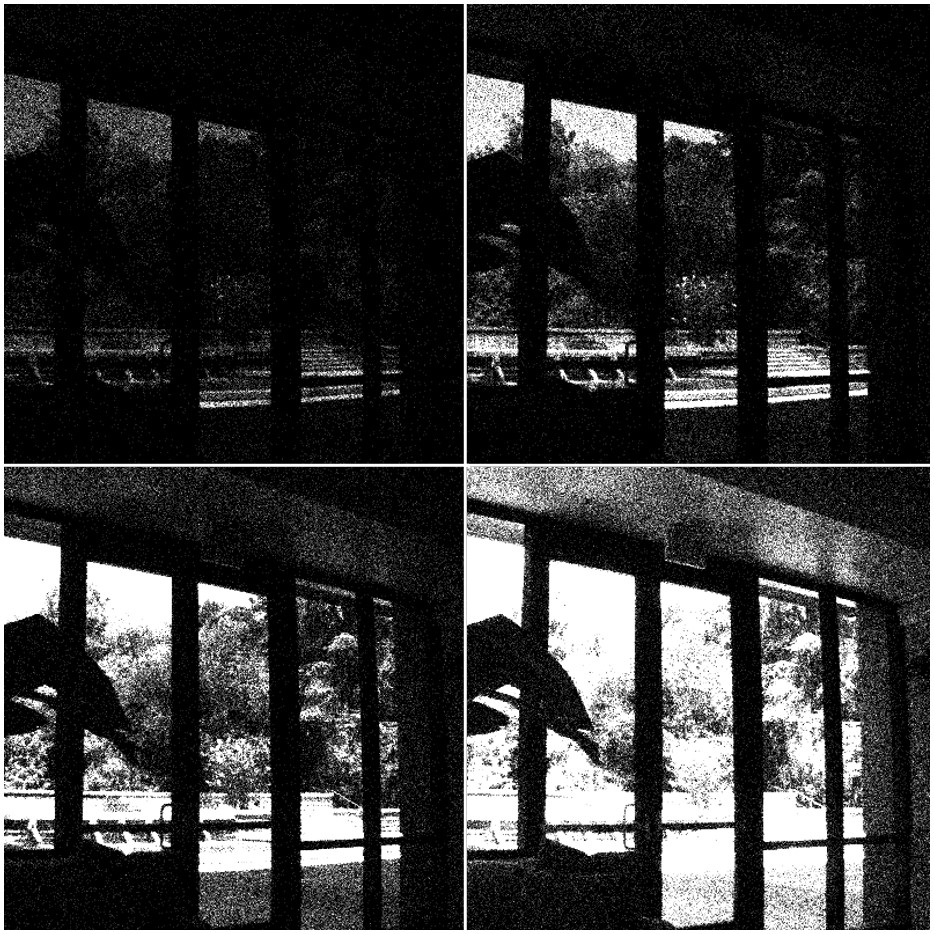}&
\hspace{-2.7ex} \includegraphics[width=0.32\linewidth]{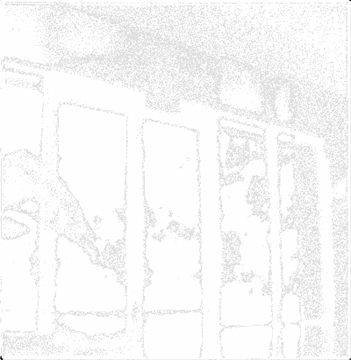}&
\hspace{-2.7ex} \includegraphics[width=0.32\linewidth]{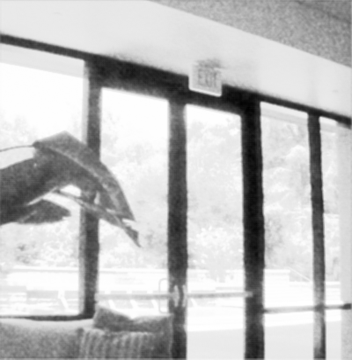}\\
\hspace{0ex} (a) Exposures &
\hspace{-2.7ex} (b) Kalantari \cite{kalantari2017deep} &
\hspace{-3ex} (c) Wu \cite{wu2018deep}\\
\end{tabular}
\caption{Existing HDR methods for spatially varying SNR. [Top] When the inputs are clean, existing HDR methods perform well. [Bottom] When the inputs are noisy, the methods fail even if they are fine-tuned using the noisy data.}
\label{fig: Figure01}
\end{figure}

\textbf{Impact of noise on HDR fusion.} The first ablation study is shown in \fref{fig: Figure01} which contains two sets of images simulated using our realistic image sensor model. The first set of images is \emph{clean} inputs simulated at 100 photons per pixel on average. Two HDR fusion algorithms \cite{kalantari2017deep, wu2018deep} were applied to blend the frames. The results are reasonable despite some minor tone differences. The second set of images is simulated at 10 photons per pixel, which is substantially noisier than the first case. The result is very bad as we can see from the figure. It also confirms that when everything remains the same, the strength of the noise will have an immediate impact on the fusion algorithm. 

\begin{figure}[h]
\centering
\includegraphics[width=\linewidth]{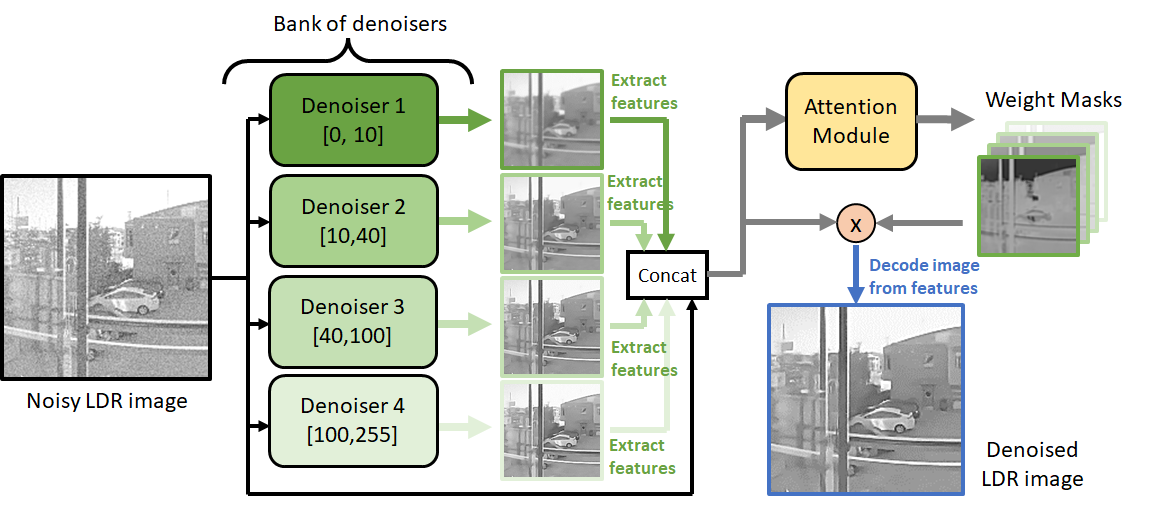}
\caption{A toy denoising (system) that uses multiple denoisers to handle the luminance range. Note that this system handles only \emph{one} exposure input. For an exposure bracket containing three or more exposures, each exposure would require one such denoising system. To alleviate the issue of requiring many denoisers, an \emph{alternative} design is presented in this paper.}
\label{fig: Figure13 Attention}
\end{figure}

\textbf{Impact of dynamic range on denoising.} The second ablation study aims to verify the impact of the dynamic range on a denoiser. To this end, we build a toy denoiser that contains four sub-denoisers as shown in \fref{fig: Figure13 Attention}. The denoisers we use are REDNet models \cite{mao2016image}. In this toy design, we send the input image to the four sub-denoisers to handle the four different luminance ranges. The denoised images are then concatenated and sent to an attention module. Five weight masks are generated to combine the denoised images and the input to form the final image. We remark that this four-denoiser system is used to handle just \emph{one} exposure in an exposure bracket. For a bracket that contains three exposures, we need three denoising systems followed by an HDR fusion module.

\begin{figure}[h]
\begin{tabular}{cc}
\multicolumn{2}{c}{\hspace{-2.4ex} \includegraphics[width=\linewidth]{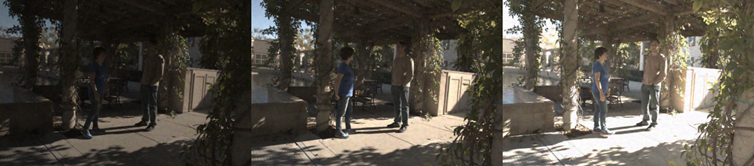}}\\
\hspace{-2.4ex} \includegraphics[width=0.49\linewidth]{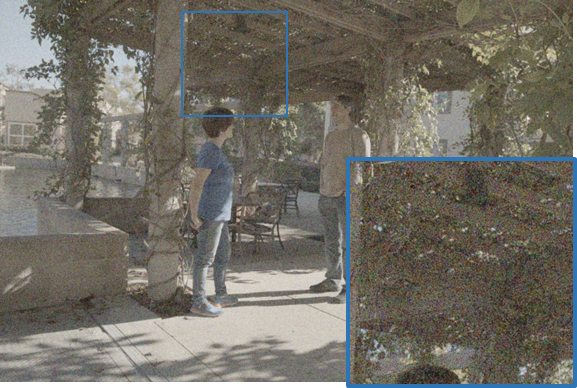}&
\hspace{-2.3ex} \includegraphics[width=0.49\linewidth]{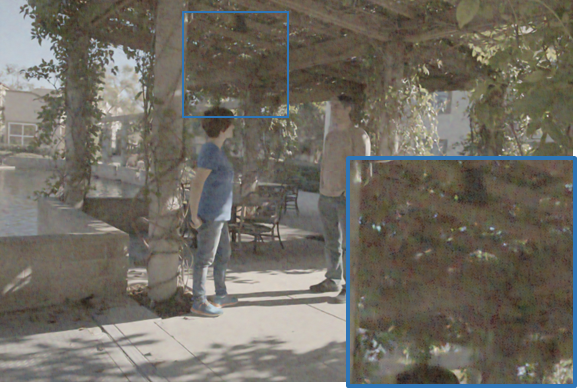}\\
\end{tabular}
\caption{HDR fusion of noisy inputs. (a) Using 3 denoisers and the vanilla fusion algorithm by \cite{kalantari2017deep}. (b) Using the denoiser in \fref{fig: Figure13 Attention}, followed by \cite{kalantari2017deep}. }
\label{fig: Figure14}
\end{figure}

\fref{fig: Figure14} shows a comparison between using this (gigantic) 12-denoiser system (4 denoisers per exposure for 3 exposures) and a 3-denoiser system (1 denoiser per exposure for 3 exposures.) As we can see in \fref{fig: Figure14}, if we have divided the exposure into a fine-enough sub-dynamic range, the denoiser is able to perform the required task. Therefore, we confirm that it is the dynamic range that limits the performance of the denoiser.

\begin{figure*}[!]
\centering
\includegraphics[width=\linewidth]{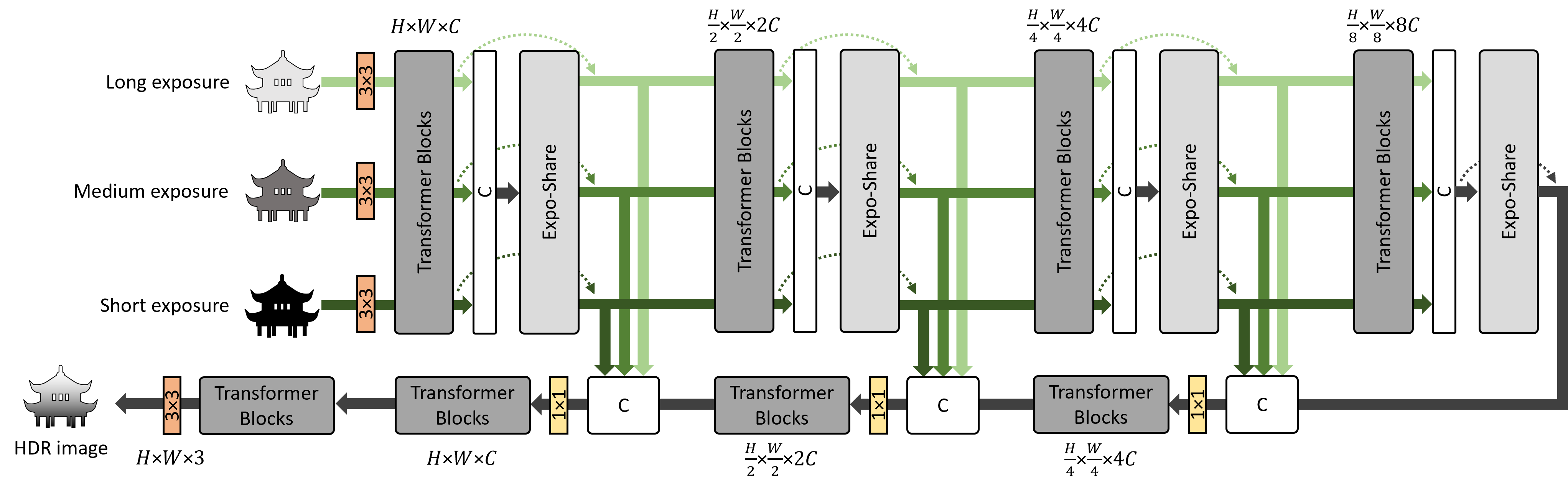}
\caption{Structure of the proposed SV-HDR network. SV-HDR has a multi-scale design, spatially adaptive multi-exposure transformer blocks to extract and fuse features, and Expo-Share blocks to exchange and blend features across exposures.}
\label{fig: Figure03 transformer}
\end{figure*}

\section{Spatially Varying HDR (SV-HDR) Network}
In this section, we present the proposed algorithm to simultaneously denoise and perform the multi-exposure fusion. Our goal is to develop a computationally efficient neural network that achieves the goal of \fref{fig: Figure13 Attention} while being much simpler. The overall architecture is shown in \fref{fig: Figure03 transformer}.

\subsection{Network architecture}
The proposed SV-HDR is an encoder-decoder architecture following the U-shape structure \cite{ronneberger2015u}. Given a set of low dynamic range (LDR) images, the encoder extracts features from each of them, whereas the decoder generates the HDR results. Previous HDR fusion methods use different branches to process short, medium, and long exposure. In SV-HDR, we propose a single encoder to process \emph{all} exposures to reduce the network capacity when handling a spatially varying SNR scene $\theta(\vx)$. This is achieved by using self-attention in the transformer block, which distinguishes the difference in exposure, gain, and noise variance. All LDR images are processed individually with the same encoder. In addition, we propose a new Expo-Share module to allow information exchange among the exposures. At the end of the encoder, three sets of features are concatenated. The decoder merges them and performs the HDR fusion.

\textbf{Multi-scale}. SV-HDR adopts a multi-scale design as shown in \ref{fig: Figure03 transformer}. The operation goes as follows. Given three LDR images $\mathbf{I}_i \in \R^{H \times W \times 3}$ at different exposures, the images are first mapped to the HDR scale by an inverse gamma correction and exposure normalization $\hat{\mathbf{I}}_i = \frac{\mathbf{I}_i^\gamma}{t_i}$, 
where $t_i$ is the exposure time factor representing the multiplicative factor of the exposures $\theta(\vx) + \mu_{\text{dark}}$. $\hat{\mathbf{I}}_i$ is then appended to $\mathbf{I}_i$ as additional channels, forming $\mathbf{J}_i \in \R^{H \times W \times 6}$ as the input to SV-HDR. SV-HDR first extracts shallow features from $\mathbf{J}_i$ by convolution, then extracts the deep features by a series of transformer blocks. The encoder uses an Expo-Share block at the end of each level to aid the feature exchange and motion mitigation. Details of the transformer blocks and Expo-Share blocks are discussed in the following sections. Down-sampling between the levels in the encoder is done by a $3 \times 3$ convolution and a pixel-unshuffling layer \cite{shi2016real}, symmetrically by convolutional and pixel-shuffling operations for up-sampling in the decoder. Skip connections are used to assist the information flow from the encoder to the decoder. Finally, the transformer blocks and a $3 \times 3$ convolution reconstruct the output HDR image $\mathbf{\hat{H}} \in \R^{H \times W \times 3}$ from the features. In our implementation, we set the number of multi-scale levels to 4.

\textbf{Loss function}. HDR imaging is about aggregating information from a wide range of signal levels. Standard loss function such as mean squared error is biased toward the bright pixels. In dark regions, subtle variations of pixel value will not be reflected in such losses but can be significantly influential compared to the signal level. Therefore, we need to compute the loss function between the tonemapped HDR image $\mathcal{T}(\mathbf{\hat{H}})$ and tonemapped ground truth HDR $\mathcal{T}(\mathbf{H})$. We use the differentiable tonemapping following \cite{kalantari2017deep, wu2018deep}.
\begin{align}
\mathcal{T}(\mathbf{H}) = \frac{\text{log}(1 + \mu \mathbf{H})}{\text{log}(1 + \mu)},
\label{eq: tonemapping}
\end{align}
where we set $\mu$ to be 5000. Our final loss function is:
\begin{align}
\mathcal{L} = \| \mathcal{T}(\mathbf{\hat{H}}) - \mathcal{T}(\mathbf{H}) \|_2 .
\label{eq: loss}
\end{align}

\subsection{Transformer Block}
We use the visual transformers as the basic components of SV-HDR. A typical transformer block consists of a multi-head self-attention (MSA) module and a feed-forward module. Among several candidates of MSA and feed-forward, we experimentally found the best performing transformer block is a Shifted Window-based MSA (SW-MSA) proposed in \cite{liang2021swinir} followed by a multi-layer perception (MLP) of two fully connected layers with a GELU activation in between. We also use residual connections to skip over each of the SW-MSA and MLP modules. The experiment details are shown in Section \ref{ablation} with results in Table \ref{tab: transformer ablation}.

SV-HDR has a serial of transformer blocks at each down-sample level. We use 4, 6, 6, and 8 transformer blocks for the four levels with down-sample ratios $1\times$, $2\times$, $4\times$, and $8\times$, respectively. The numbers of heads of MSA are 1, 2, 4, and 8 in the four levels. In the final refinement stage, four additional transformer blocks are used. The number of channels of features, $C$, is selected to be 48.

\subsection{Expo-Share Block}
\begin{figure}[h]
\centering
\includegraphics[width=\linewidth]{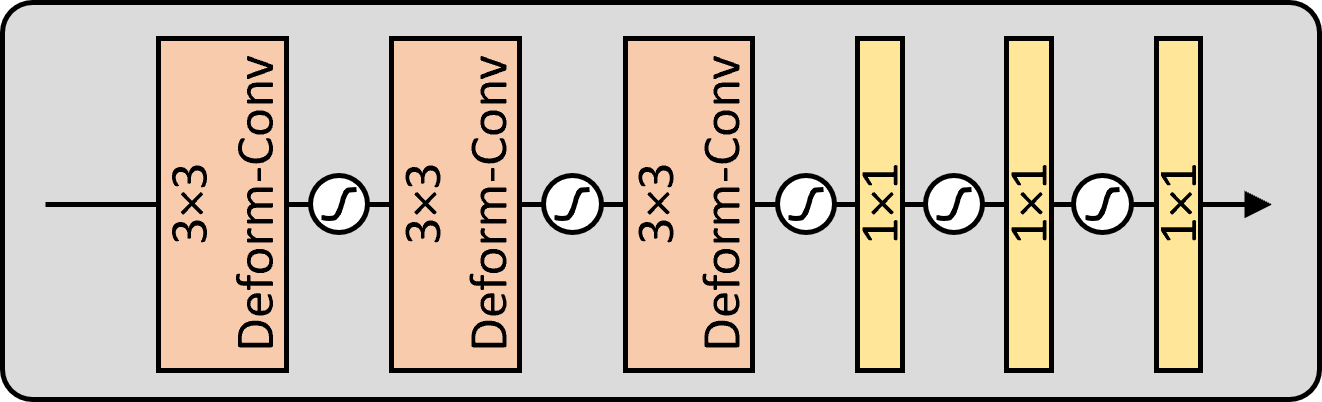}
\caption{The proposed Expo-Share Block. Each block contains three sections of deformable convolutions and three $1 \time 1$ convolutions. The activation functions are the GELUs.}
\label{fig: Figure03 ExpoShare}
\end{figure}

To promote temporal information sharing across the exposures, we design an Expo-Share Block that jointly processes features from all exposures at the end of each level. At level $l$, the features $\mathbf{F}_i \in \R^{\frac{H}{2^l} \times \frac{W}{2^l} \times 2^lC}$ are concatenated in the channel dimension before Expo-Share and are split after it, denoted $\Tilde{\mathbf{F}}_i \in \R^{\frac{H}{2^l} \times \frac{W}{2^l} \times 2^lC}$. Considering that the scene can be dynamic across the frames, we use deformable convolutions \cite{dai2017deformable, zhu2019deformable} to augment the spatial sampling region and implicitly align the features. Each Expo-Share Block contains three $3 \times 3$ deformable convolution layers, followed by three $1 \times 1$ convolutions to aggregate pixel-wise information across channels. GELU activation is used between each two consecutive layers. The overall operation of an Expo-Share block is shown in \fref{fig: Figure03 ExpoShare}. Features processed by the Expo-Share blocks are added back to itself with residual connections $\Tilde{\mathbf{F}}_i + \mathbf{F}_i$ as illustrated in \fref{fig: Figure03 transformer}, and go through the remaining encoder computations individually. 

\begin{figure*}[t]
    \captionsetup[subfloat]{font=scriptsize}
    \centering
    \includegraphics[width=\linewidth]{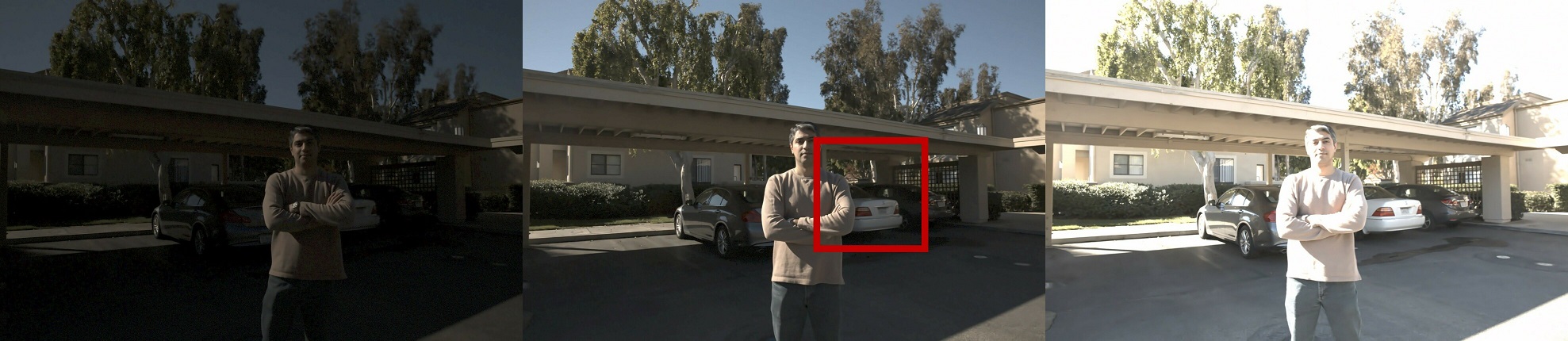}\\ 
   \vspace{-0.6ex}
    \subfloat[Input / \textbf{0.644 lux}]{%
    \includegraphics[width=0.142\linewidth]{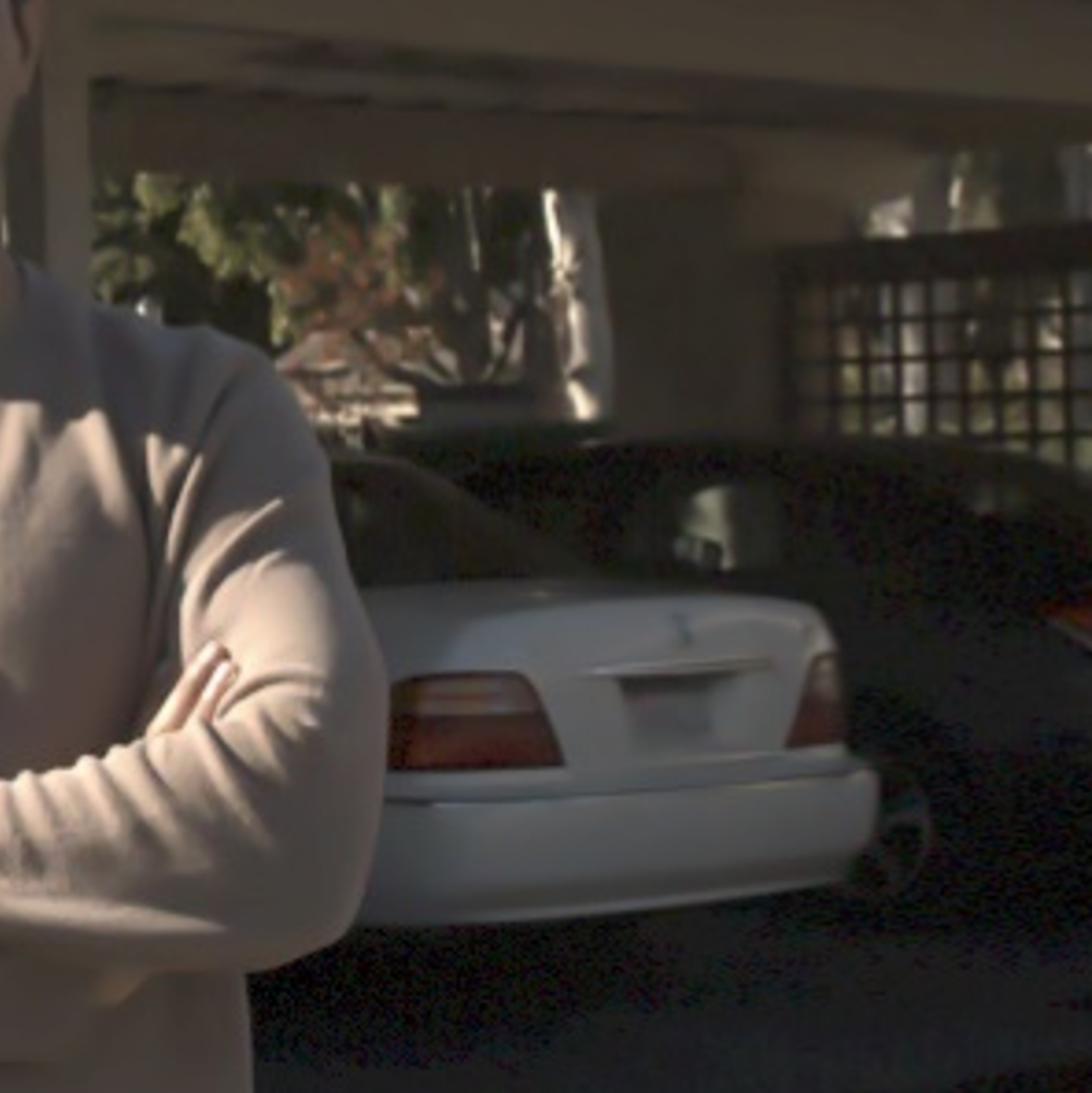}}
    \hfill
    \subfloat[Kalantari \cite{kalantari2017deep}]{%
    \includegraphics[width=0.142\linewidth]{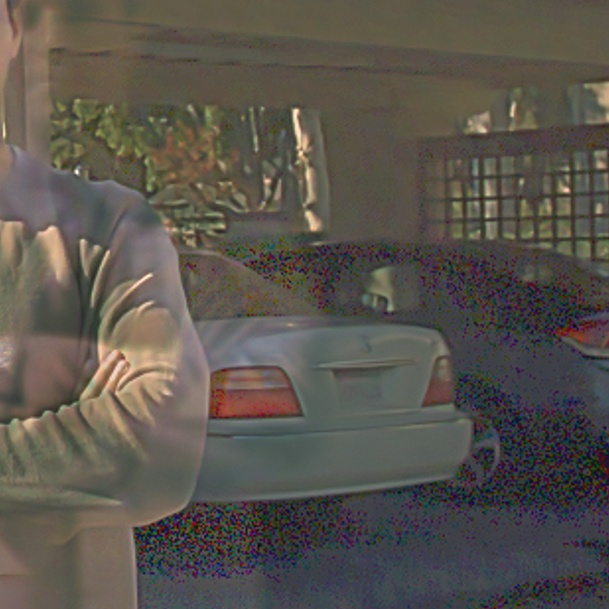}}
   \hfill
    \subfloat[Wu \cite{wu2018deep}]{%
    \includegraphics[width=0.142\linewidth]{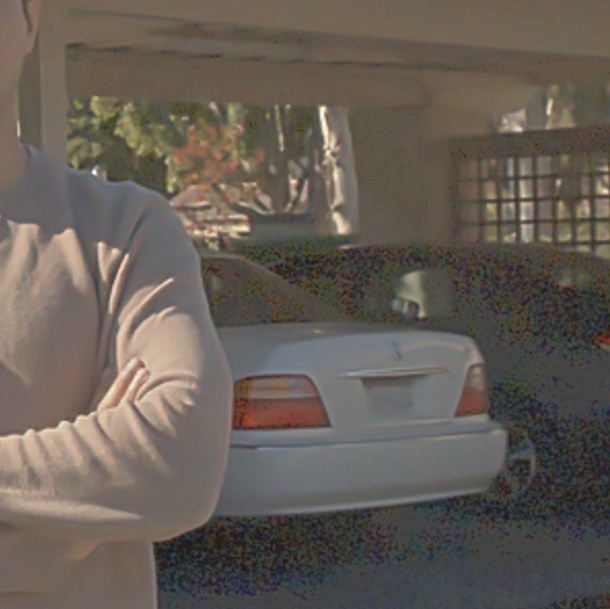}}
     \hfill
    \subfloat[AHDRNet \cite{yan2019attention}]{%
    \includegraphics[width=0.142\linewidth]{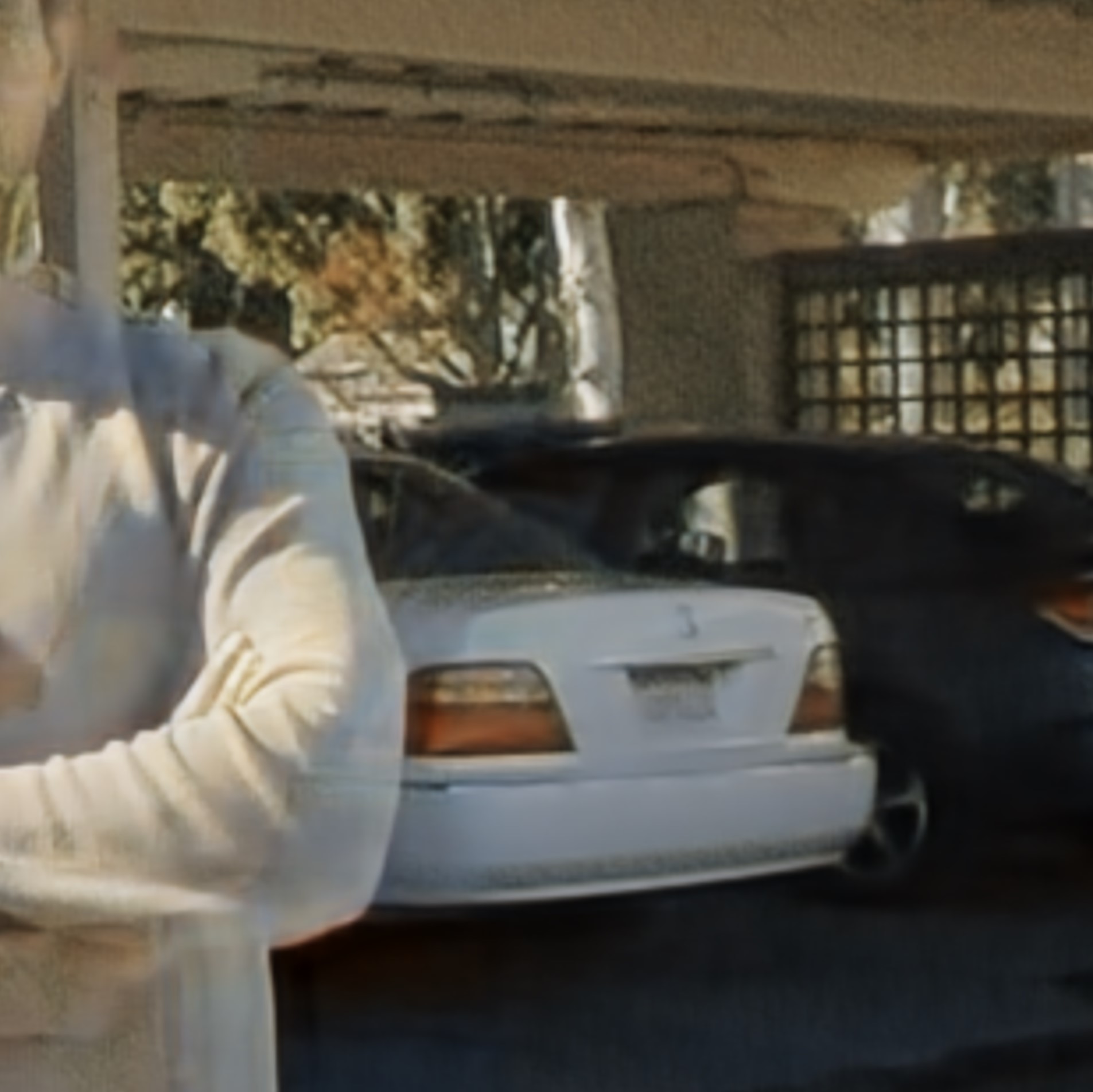}}
   \hfill
  \subfloat[NHDRRNet \cite{yan2020deep}]{%
    \includegraphics[width=0.142\linewidth]{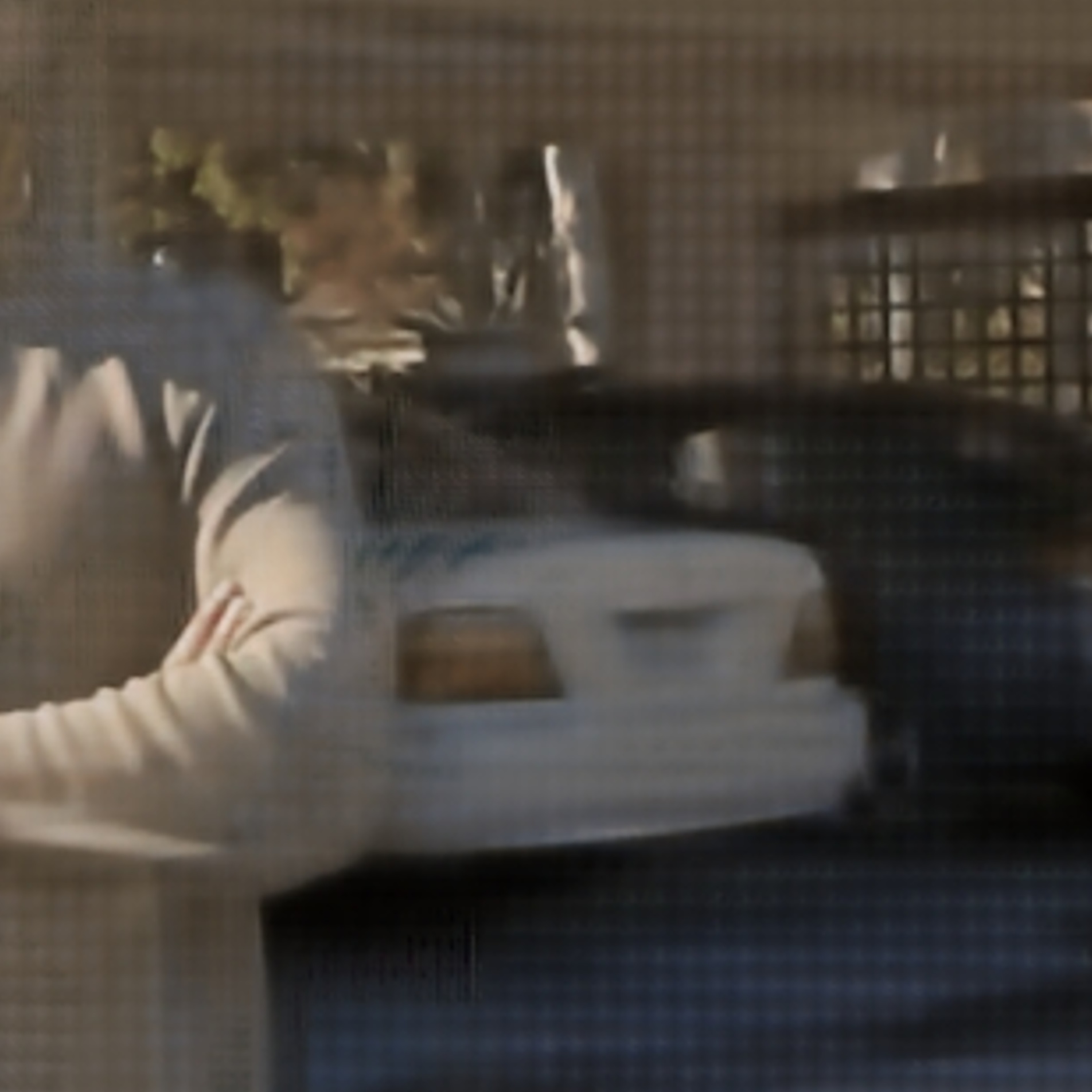}}
   \hfill
  \subfloat[\textbf{SV-HDR [Ours]}]{%
    \includegraphics[width=0.142\linewidth]{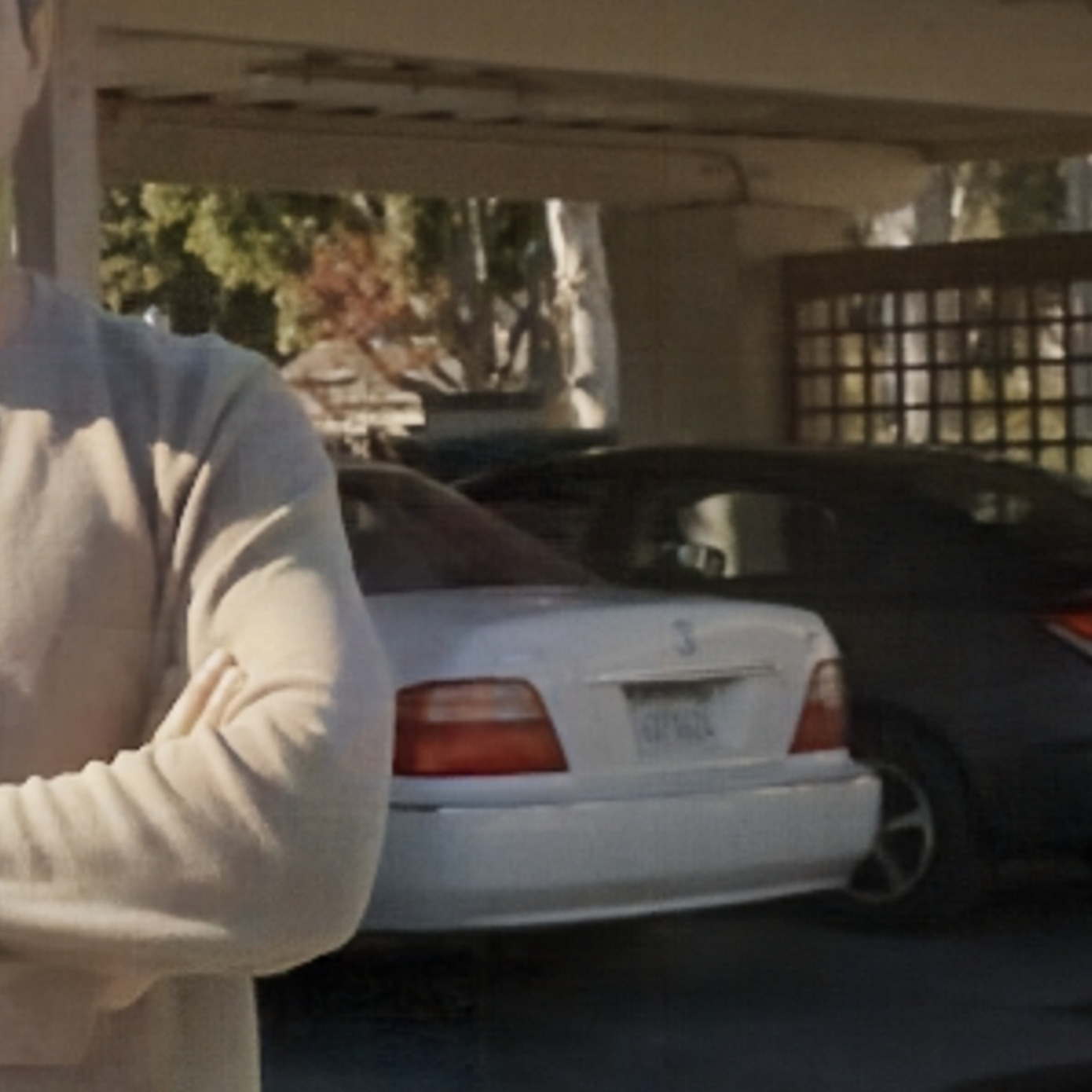}}
   \hfill
  \subfloat[Ground Truth]{%
    \includegraphics[width=0.142\linewidth]{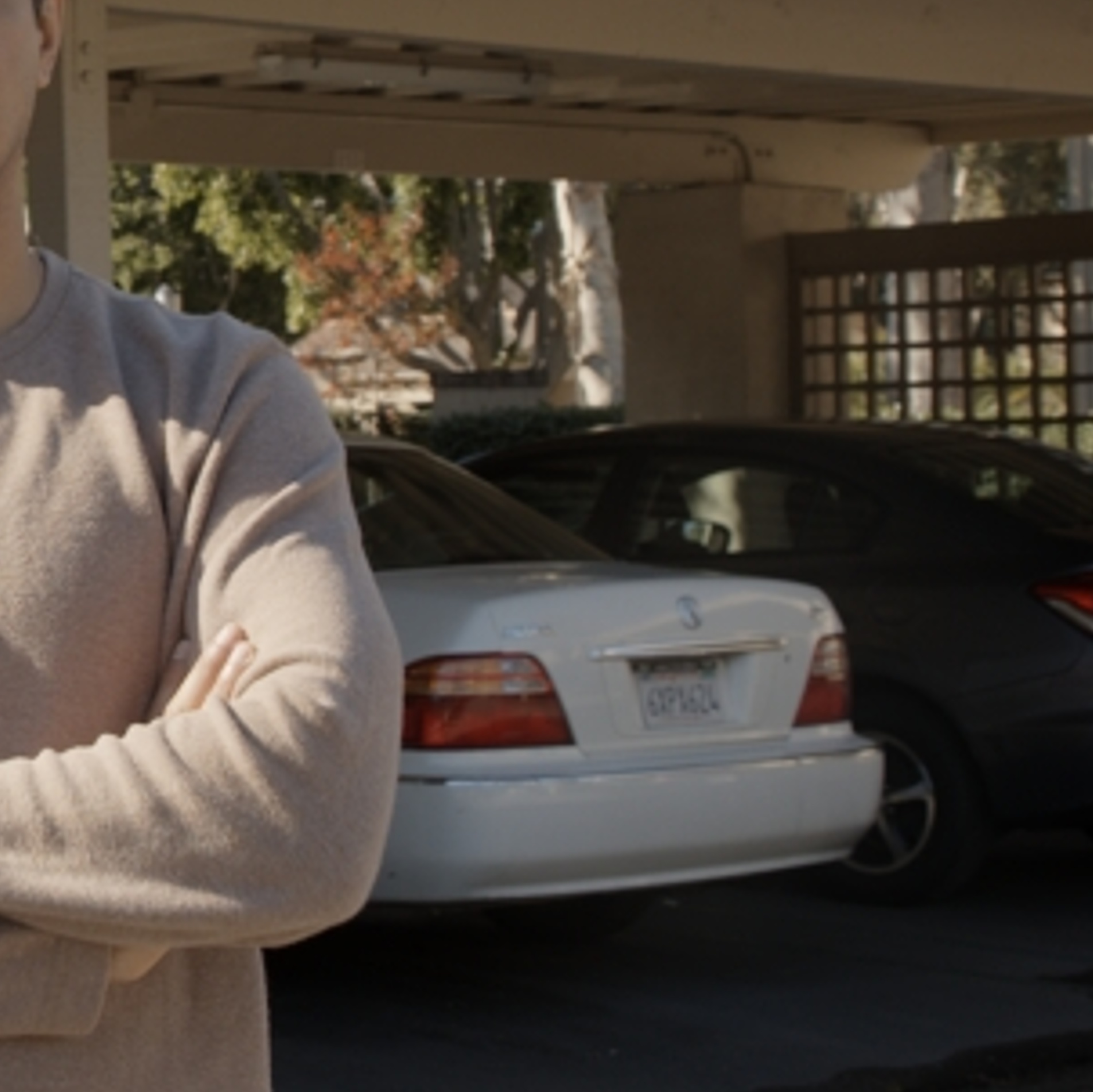}}
    \\
    \vspace{1ex}
  \includegraphics[width=\linewidth]{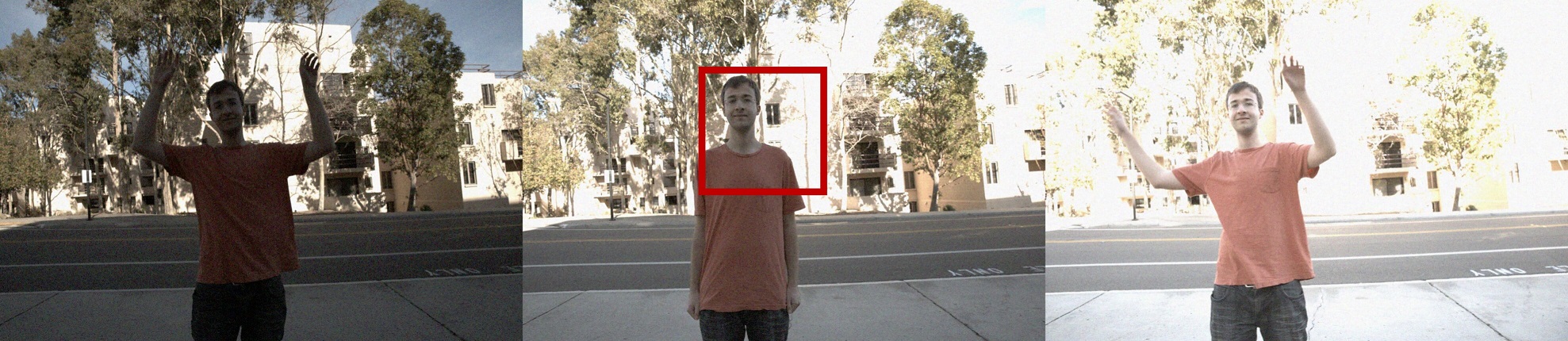} \\
  \vspace{-0.3ex}
    \subfloat[Input / \textbf{0.080 lux}]{%
    \includegraphics[width=0.142\linewidth]{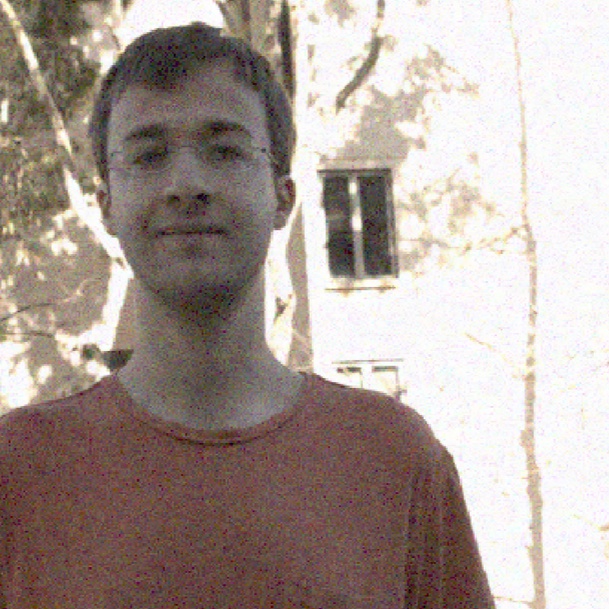}}
    \hfill
    \subfloat[Kalantari \cite{kalantari2017deep}]{%
    \includegraphics[width=0.142\linewidth]{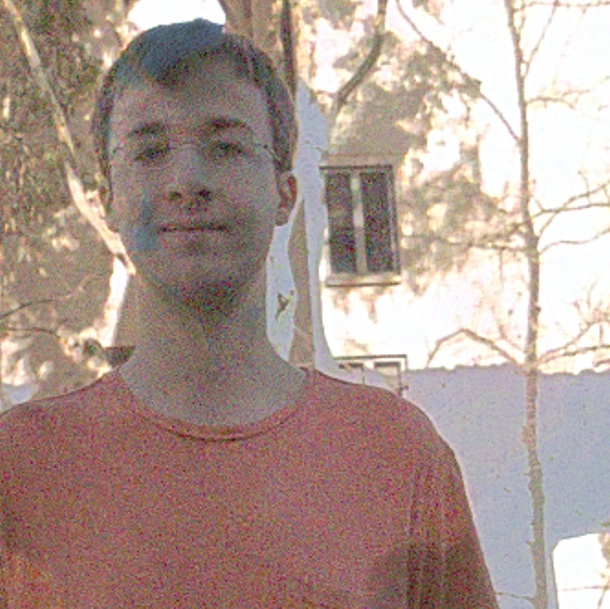}}
   \hfill
    \subfloat[Wu \cite{wu2018deep}]{%
    \includegraphics[width=0.142\linewidth]{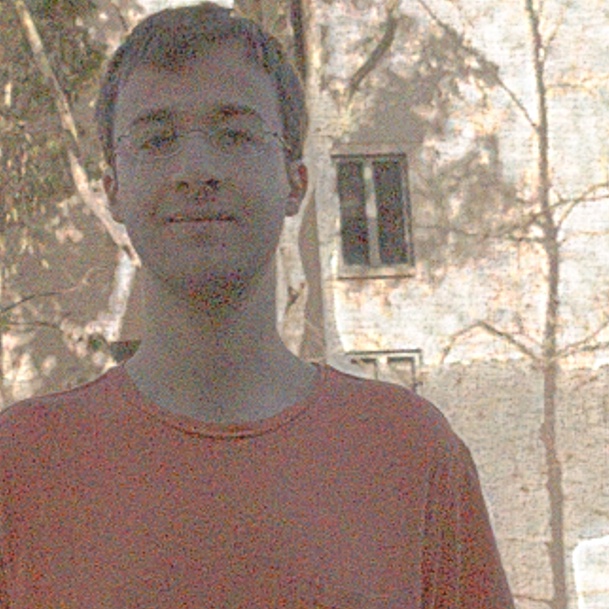}}
    \hfill
    \subfloat[AHDRNet \cite{yan2019attention}]{%
    \includegraphics[width=0.142\linewidth]{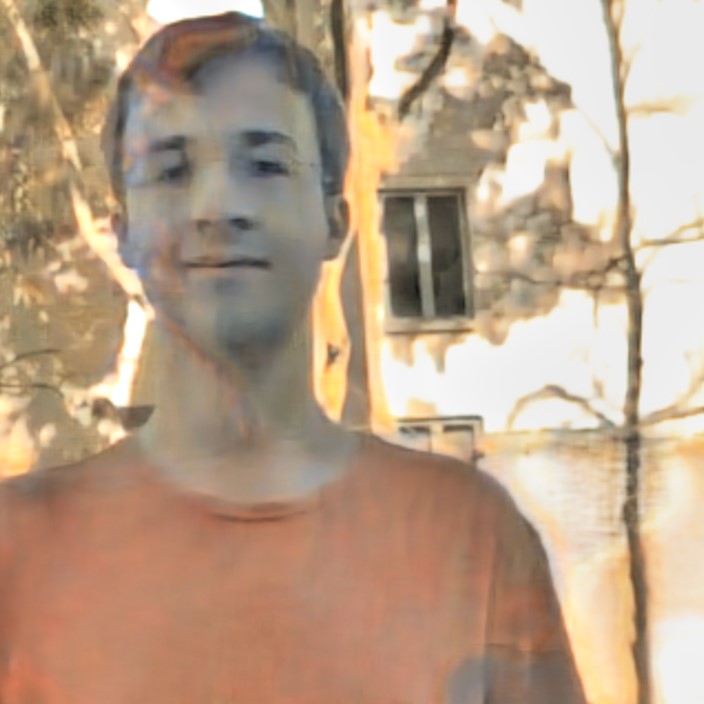}}
 \hfill
  \subfloat[NHDRRNet \cite{yan2020deep}]{%
    \includegraphics[width=0.142\linewidth]{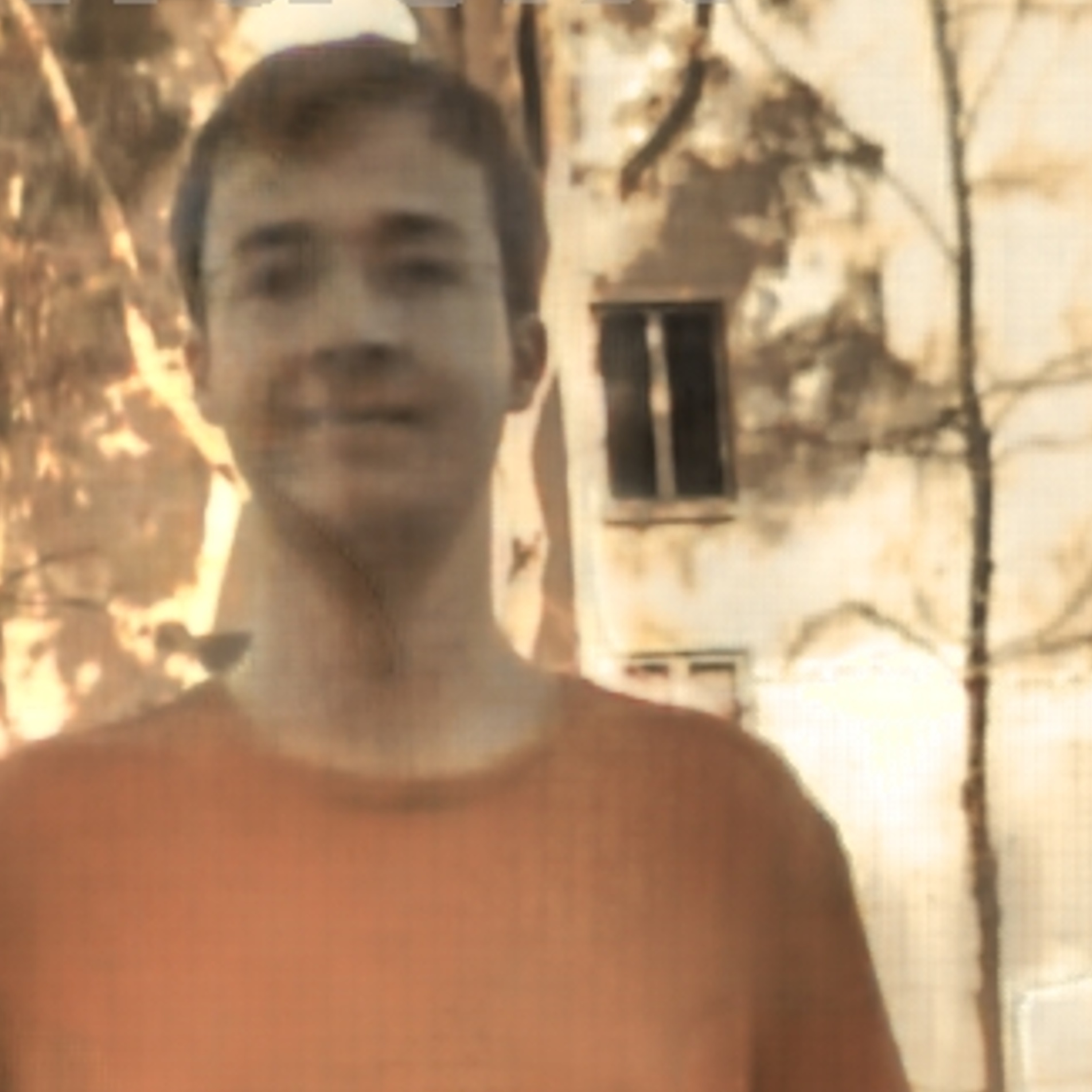}}
   \hfill
  \subfloat[\textbf{SV-HDR [Ours]}]{%
    \includegraphics[width=0.142\linewidth]{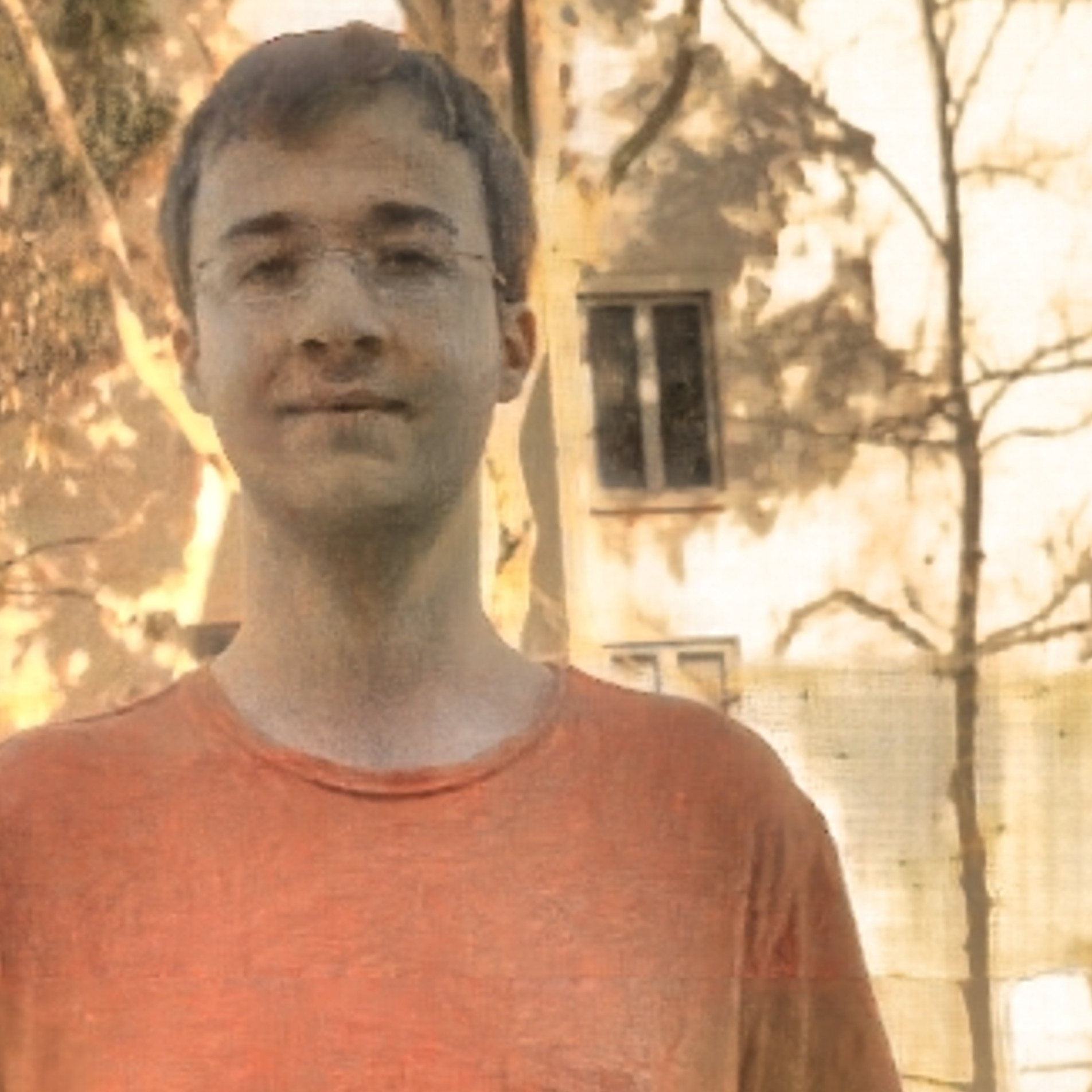}}
   \hfill
  \subfloat[Ground Truth]{%
    \includegraphics[width=0.142\linewidth]{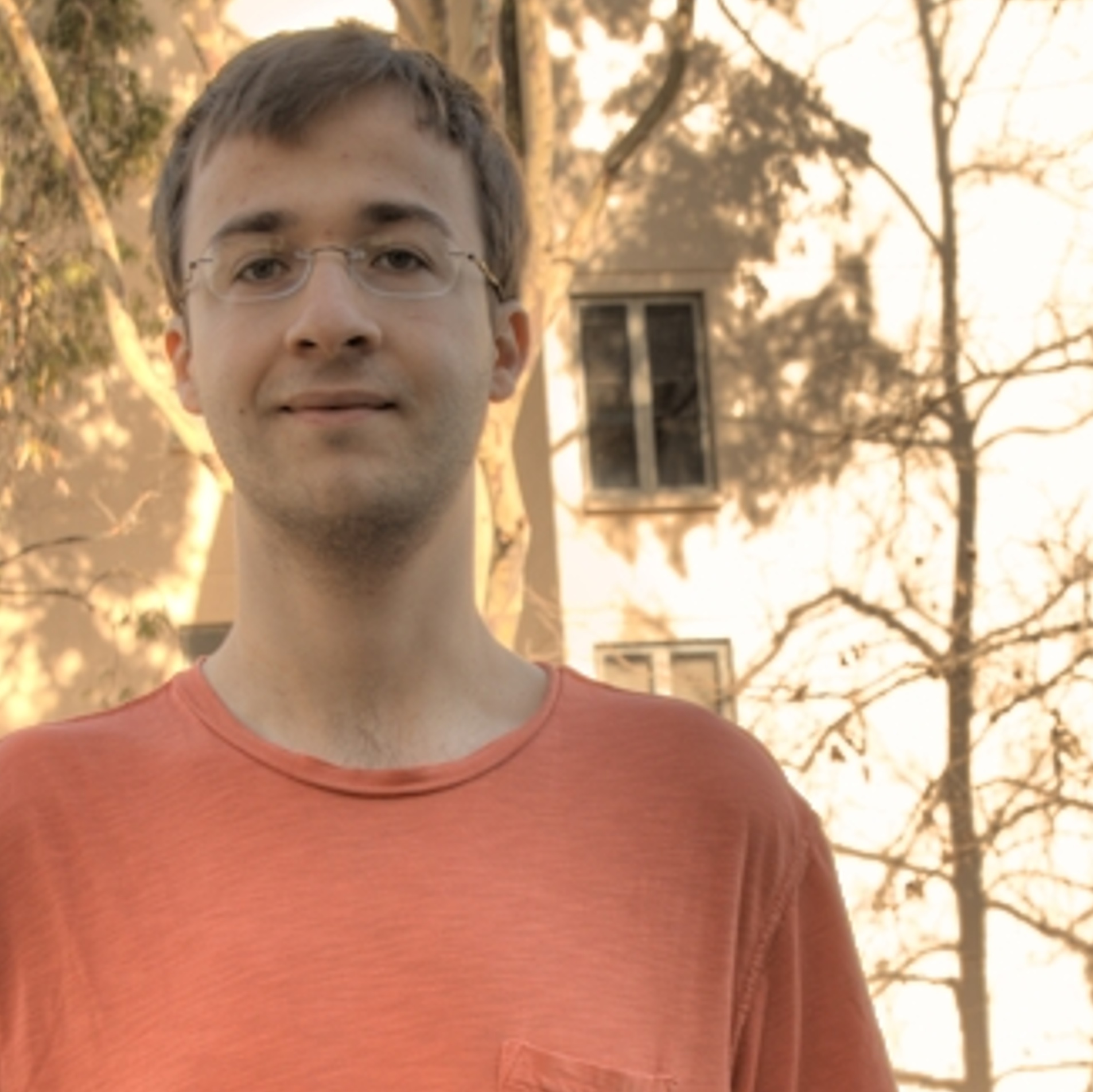}}\\
  \caption{Qualitative comparison using synthetic data. The input images of this figure are synthesized using the realistic sensor model and parameters. The corresponding photon level is marked in the input image. Dataset: \cite{kalantari2017deep}.}
  \vspace{-1ex}
  \label{fig: Figure17 Synthetic Experiment}
\end{figure*}

\vspace{-3ex}
\section{Experiments}
\subsection{Implementation Details}
We train our model using the Kalantari dataset \cite{kalantari2017deep}. This dataset contains a large number of LDR images with a ground truth HDR image aligned with the middle exposure. We first augment the data by 8 times by randomly rotating and flipping the images, and we down-sample the images by $2\times$ and randomly crop the centers of images to $128 \times 128$ patches, where the down-sampling and cropping from center ensure these patches contain dynamic foreground objects. We then synthesize the noisy LDR image following our sensor model. To make our model applicable to a wide range of lighting conditions, we randomly sample $\tau$ such that the maximum photon count per pixel has a triangular distribution between 4 and 256 with a mode of 8. This is roughly equivalent to 0.005 to 0.323 lux at a typical exposure time of 1/50 sec and a pixel pitch of 6 microns for light with 400 to 700 nm wavelengths \cite{Palum2002HowMP}. We fix QE at 50\% and $\alpha$ at 1. The read noise $\sigma_{\text{read}}$ is set to 0.0292, 0.1798, 1.4384 for short, medium, and long exposures, respectively, matching the characteristics of a Sony ILCE-7M2 camera at ISO of 200, 1600, and 12800 to enable testing on real images.

We train our model 1184 iterations per epoch for 300 epochs with a batch size of 3. We use the loss function in Eq. \ref{eq: loss}. We use the Adam optimizer \cite{kingma2014adam} with parameters $\beta_1$ = 0.9 and $\beta_2$ = 0.999. Our learning rate is scheduled by a cosine scheduler with an initial rate of $0.0001$. The training of our model takes 32 hours using an RTX 2080 Ti GPU.

\subsection{Experiment Setup}
We test our model on the test set of \cite{kalantari2017deep}. The noise synthesis process again follows \ref{eq: main sensor model}. We fix $\tau$ at specific levels to conduct the qualitative and quantitative comparisons. We plot the test performance as curves PSNR versus $\tau$.

We also conduct denoising and HDR fusion experiments on real images. We capture images with a Sony ILCE-7M2 camera and a Sigma Art 24-70mm F2.8 DG DN lens. We conduct two sets of real experiments. (I) We capture three images at a fixed ISO of 3200 and an aperture of f/2.8. The exposure time is 1/1250 sec, 1/160 sec, and 1/20 sec for the short, medium, and long exposure images. The setup of this experiment ensures the Poisson shot noise characteristics match our training setup. We further conduct a generalization experiment. (II) At each scene, we capture three images at a fixed exposure of 1/50 second, an aperture of f/5.0, and at varying ISO of 200, 1600, and 12800. Varying ISO is preferred to varying exposure time in certain imaging conditions, but a higher ISO amplifies the Poisson shot noise variance and corresponds to a higher read noise, so real experiment (II) is considered more challenging. SV-HDR can handle this challenging problem because it is adaptive to the non-uniform noise variance.

We compare the performances with Kalantari \cite{kalantari2017deep}, Wu \cite{wu2018deep}, AHDRNet \cite{yan2019attention}, NHDRRNet \cite{yan2020deep}. As we demonstrate in Section \ref{preliminary experiment}, fine-tuning Kalantari's and Wu's models under our setup do not converge, nor do their fine-tuned models generate high-fidelity HDR reconstruction. Therefore, we test their model as-is. \cite{yan2019attention} and \cite{yan2020deep} have more sophisticated designs, so we fine-tune their models for 50,000 additional iterations in 50 epochs with batches of 8 and 32 samples sized $256 \times 256$. The learning rates start at 0.0001 and are scheduled using a polynomial decay with a power of 0.9. 

\subsection{Synthetic Experiments}

\textbf{Qualitative comparisons}. We visually compare the results. \fref{fig: Figure17 Synthetic Experiment} shows the tone-mapped reconstruction results at a scene brightness roughly equivalent to 0.644 lux and 0.080 lux. Kalantari's and Wu's methods produce noisy HDR results. From \fref{fig: Figure17 Synthetic Experiment} (b), we also find that the optical flow aligning mechanism fails when noise is strong. This causes the fusion results to have ghosting effects. Although fine-tuned on the realistic sensor model, NHDRRnet cannot adapt sufficiently to the spatially varying SNR, making the fusion output over-smoothed at low-light regions. For example, the cars in the top image are over-smoothed. The proposed SV-HDR successfully removes noise and retains details at various lighting conditions.

\begin{figure}[h]
\centering
\includegraphics[width=0.95\linewidth]{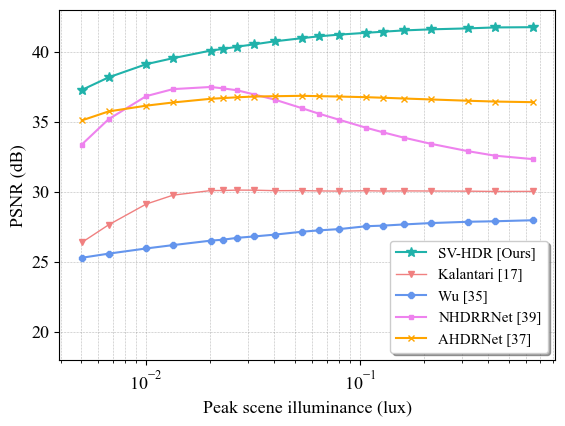}
\vspace{-1ex}
\caption{PSNR Curves of various HDR fusion algorithms. We note that SV-HDR stays at the top across all illumination levels.}
\label{fig: Figure15 PSNR Curves}
\end{figure}

\textbf{Quantitative comparisons}. We quantitatively analyze the performances using PSNR and MS-SSIM metrics. We also calculate PSNR and MS-SSIM on tonemapped images (PSNR-$\mu$ and MS-SSIM-$\mu$) following \ref{eq: tonemapping}. We set exposure $\tau$ such that the peak scene illuminance is roughly 0.4, 0.2, 0.1, and 0.05 lux. Table \ref{tab: PSNR} shows the average PSNR and MS-SSIM scores. SV-HDR outperforms all competing methods in all metrics. We consider HDR-VDP-2 \cite{HDR-VDP-2} a less suitable metric, because it is less sensitive to distortions in dark regions while our experiments simulate images captured at photon-limited scenes. We further plot and compare the PSNR curves versus the illuminance in \fref{fig: Figure15 PSNR Curves}. The proposed SV-HDR has a consistently higher PSNR for an illumination level from 0.005 lux to 0.644 lux. We observe that NHDRRnet has a PSNR peak at about 0.01 lux where the majority of training samples lie, and its PSNR quickly drops as the illuminance decreases or increases. This phenomenon also suggests the difficulty of using a single network to handle a wide range of noise variance.
\begin{table*}[h]
    \centering
    \setlength{\aboverulesep}{0pt}
    \setlength{\belowrulesep}{0pt}
    \resizebox{\textwidth}{!}{
    \begin{tabular}{l|cccc|cccc}
        \toprule[1pt]
         Illuminance (lux) & 0.4 & 0.2 & 0.1 & 0.05 & 0.4 & 0.2 & 0.1 & 0.05 \\
         \hline
         Method & \multicolumn{4}{c|}{PSNR / PSNR-$\mu$ (dB)} & \multicolumn{4}{c}{MS-SSIM / MS-SSIM-$\mu$} \\
         \hline
         Kalantari \cite{kalantari2017deep} & 30.04/20.33   & 30.08/19.40   & 30.10/18.11   & 30.11/16.53    
                                            & 0.8762/0.8661 & 0.8768/0.8202 & 0.8769/0.7540 & 0.8788/0.6749   \\
         Wu \cite{wu2018deep}               & 27.92/20.28   & 27.79/19.56   & 27.56/18.44   & 27.17/16.98    
                                            & 0.8215/0.7936 & 0.8151/0.7569 & 0.8055/0.7042 & 0.7914/0.6422   \\
         AHDRNet \cite{yan2019attention}    & 36.46/29.40   & 36.61/30.77   & 36.77/31.92   & 36.87/31.94    
                                            & 0.9664/0.9475 & 0.9687/0.9513 & 0.9711/0.9530 & 0.9721/0.9511   \\
         NHDRRNet \cite{yan2020deep}        & 32.60/22.46   & 33.45/23.78   & 34.59/25.35   & 35.99/27.24    
                                            & 0.9282/0.8481 & 0.9398/0.8784 & 0.9510/0.9050 & 0.9614/0.9231    \\
         \rowcolor{Gray}\textbf{SV-HDR [Ours]} & \textbf{41.75/37.30} & \textbf{41.61/37.38} & \textbf{41.36/36.73} & \textbf{40.98/35.70} 
                                               & \textbf{0.9835/0.9826} & \textbf{0.9831/0.9801} & \textbf{0.9822/0.9740} & \textbf{0.9805/0.9616} \\
         \bottomrule[1pt]
    \end{tabular}}
    \caption{Reconstruction quality metric comparisons of various HDR fusion algorithms on a synthetic testing dataset.}
    \label{tab: PSNR}
\end{table*}

\vspace{-2ex}

\subsection{Real Experiments}
Visual comparisons for the real experiments are shown in \fref{fig: Figure18 Real Experiment}, where the top images show real experiment (I) and the bottom showing (II). We further show gamma-corrected raw images to highlight the presence of spatially varying noise. The reconstructions in (I) resonate with those of synthetic experiments. The competing methods, however, fail to fuse the low-light foreground to their reconstruction outputs in (II). In contrast, SV-HDR recovers these details and generates a wider dynamic range.

\begin{table}[h]
    \centering
    \setlength{\aboverulesep}{0pt}
    \setlength{\belowrulesep}{0pt}
    \begin{tabular}{cc|c}
        \toprule[1pt]
         Encoder   & Decoder   & PSNR (dB) \\
         \hline
         MDTA   & W-MSA  & 25.48 \\
         MDTA   & SW-MSA & 25.48 \\
         MDTA   & MDTA   & 39.68 \\
         W-MSA  & W-MSA  & 38.91 \\
         W-MSA  & SW-MSA & 39.33 \\
         W-MSA  & MDTA   & 39.37 \\
         SW-MSA & W-MSA  & 39.39 \\
         \textbf{SW-MSA} & \textbf{SW-MSA} & \textbf{40.28} \\
         SW-MSA & MDTA   & 39.90 \\
         \bottomrule[1pt]
    \end{tabular} \\
    \vspace{2ex}
    \centering
    \begin{tabular}{l|cccc}
        \toprule[1pt]
         Feed-Forward   & LeFF & DeLeFF & GDFN & \textbf{MLP} \\
         \hline
         PSNR (dB)  & 39.65  & 39.88 & 40.09 & \textbf{40.34}\\
         \bottomrule[1pt]
    \end{tabular} 
    \caption{Ablation study of different choices of encoders and decoders, and choices of feed-forward models.}
    \label{tab: transformer ablation}
\end{table}

\begin{figure}[h!]
\centering
\includegraphics[width=0.95\linewidth]{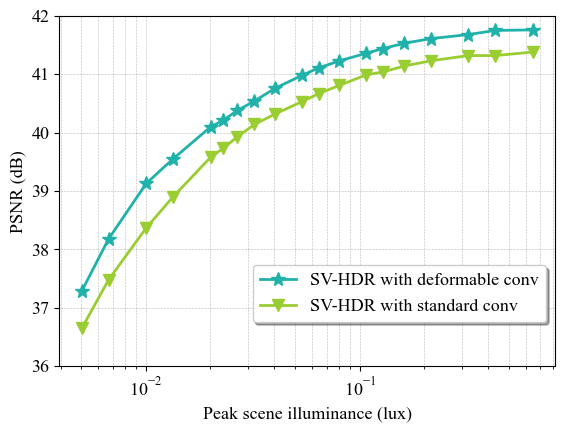}
\vspace{-1ex}
\caption{Ablation study where the deformable convolution is replaced by the standard convolution.}
\vspace{-2ex}
\label{fig: Figure15 dc ablation}
\end{figure}

\begin{figure*}[t]
    \captionsetup[subfloat]{font=scriptsize}
    \centering
    \includegraphics[width=\linewidth]{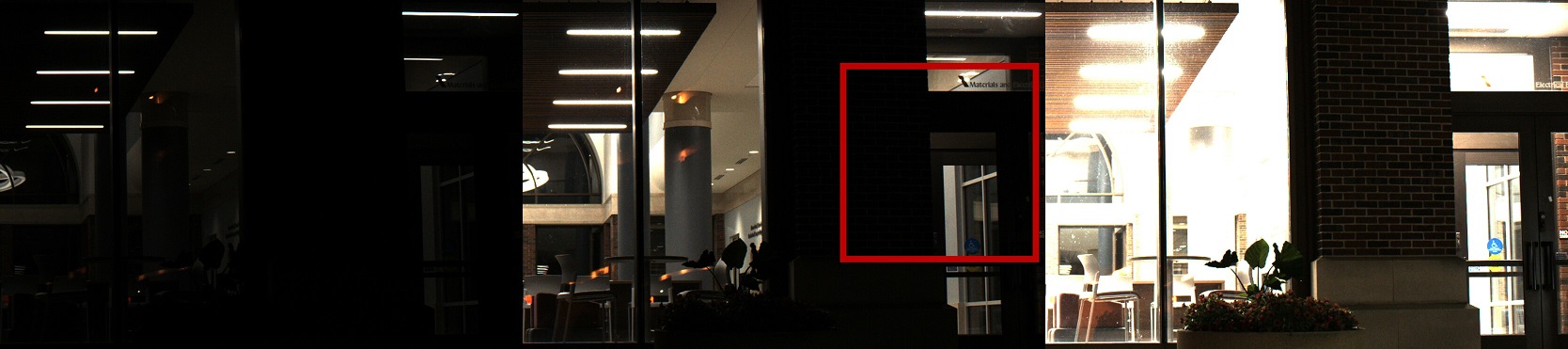}\\
    \vspace{-0.3ex}
    \subfloat[Input / \textbf{1/160 sec}]{%
    \includegraphics[width=0.165\linewidth]{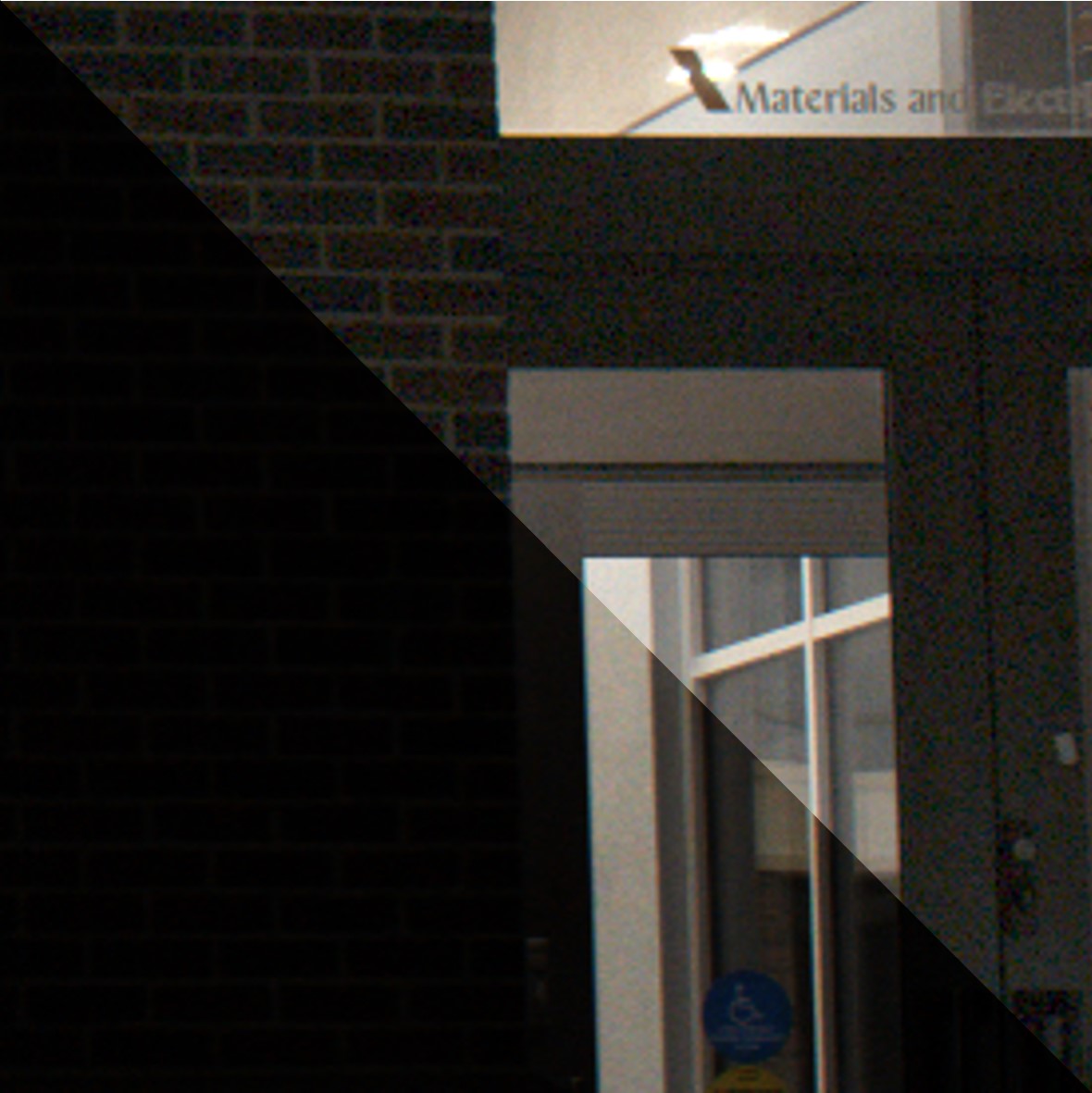}}
    \hfill
    \subfloat[Kalantari \cite{kalantari2017deep}]{%
    \includegraphics[width=0.165\linewidth]{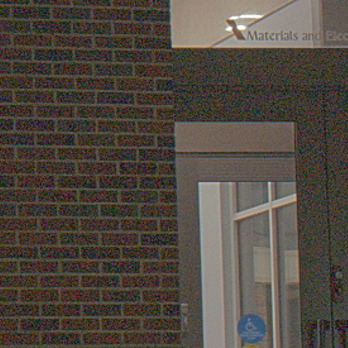}}
   \hfill
    \subfloat[Wu \cite{wu2018deep}]{%
    \includegraphics[width=0.165\linewidth]{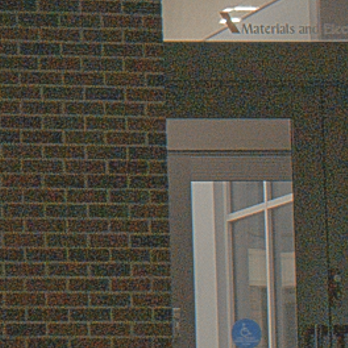}}
    \hfill
    \subfloat[AHDRNet \cite{yan2019attention}]{%
    \includegraphics[width=0.165\linewidth]{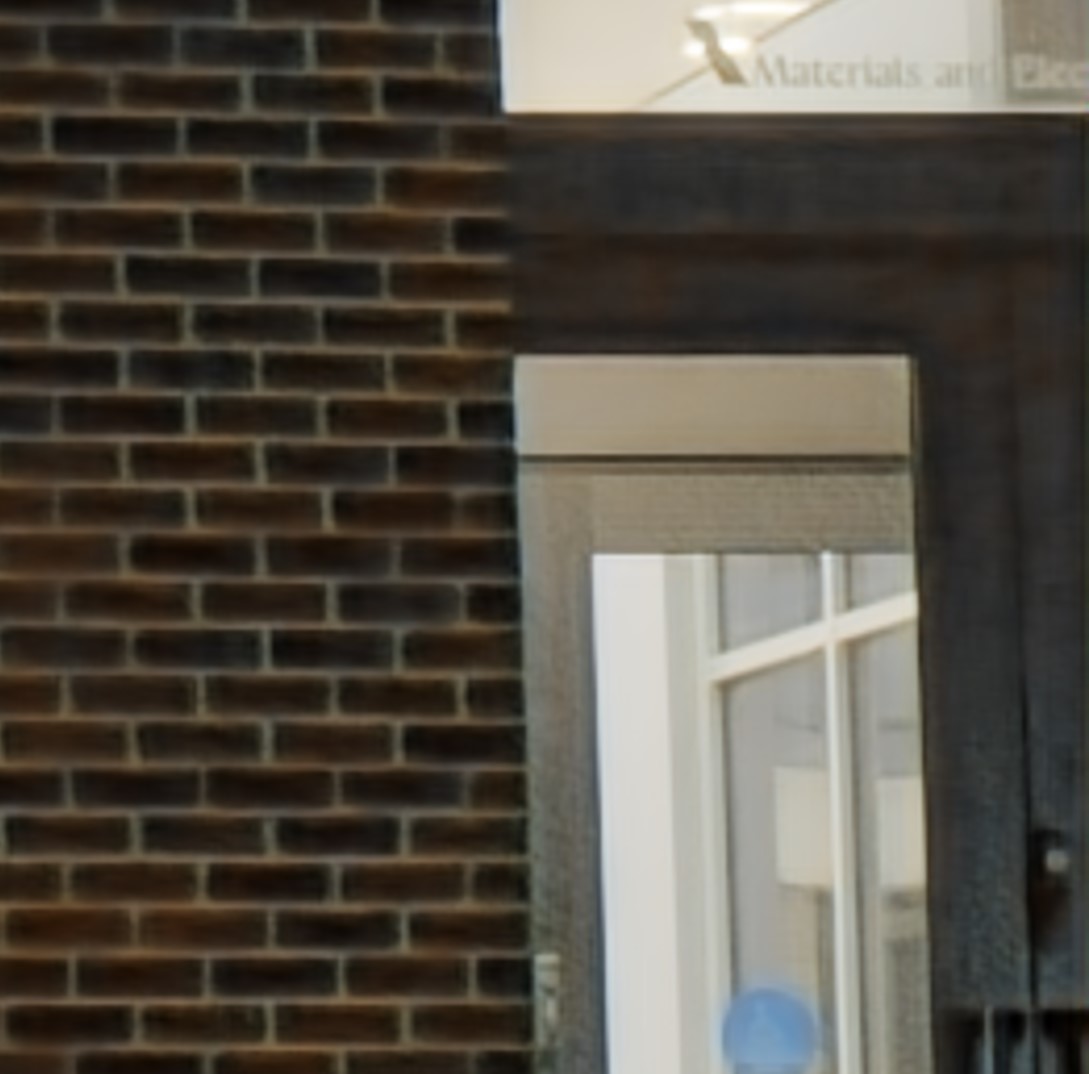}}
 \hfill
  \subfloat[NHDRRNet \cite{yan2020deep}]{%
    \includegraphics[width=0.165\linewidth]{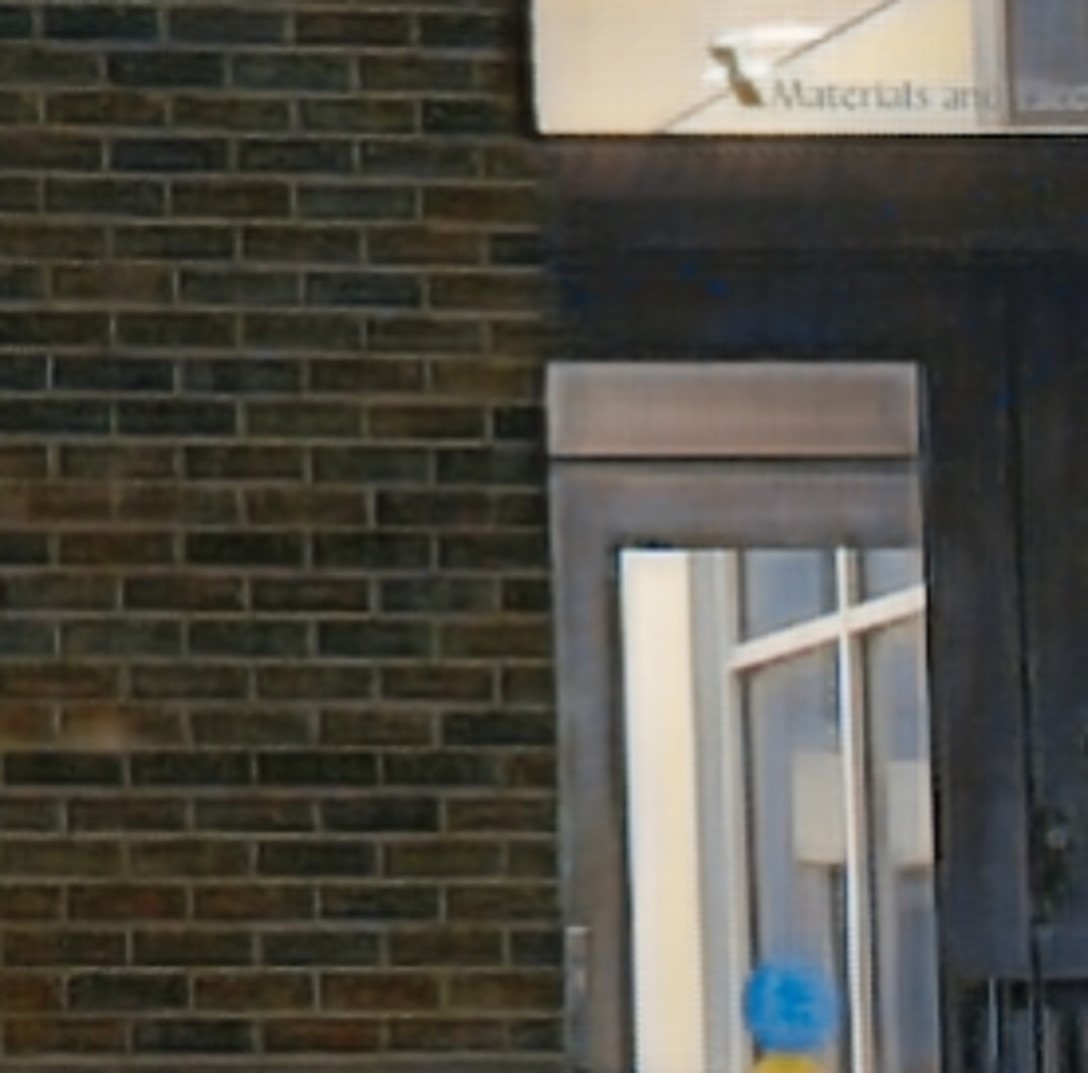}}
   \hfill
  \subfloat[\textbf{SV-HDR [Ours]}]{%
    \includegraphics[width=0.165\linewidth]{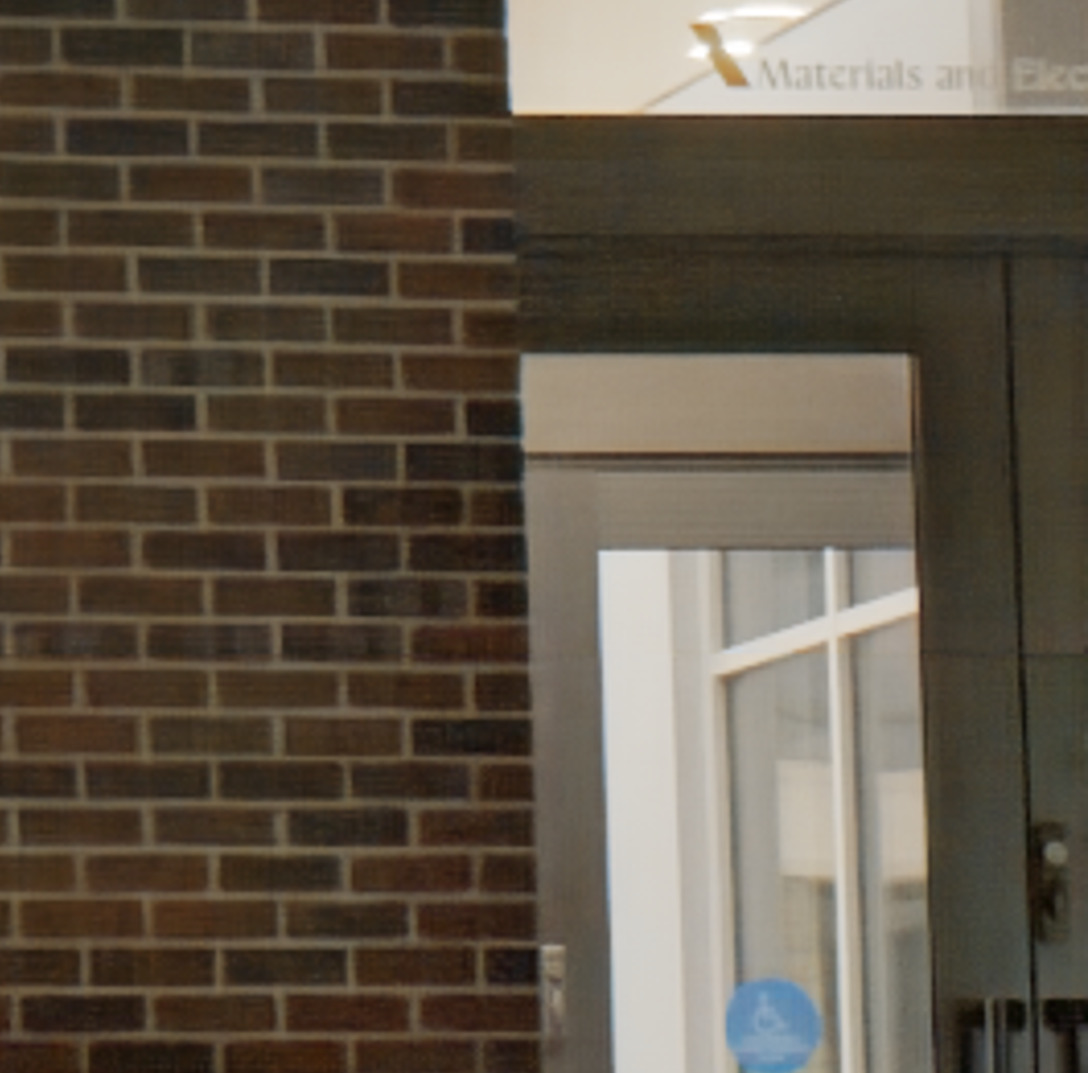}}
    \\
    \vspace{1ex}
    
  \includegraphics[width=\linewidth]{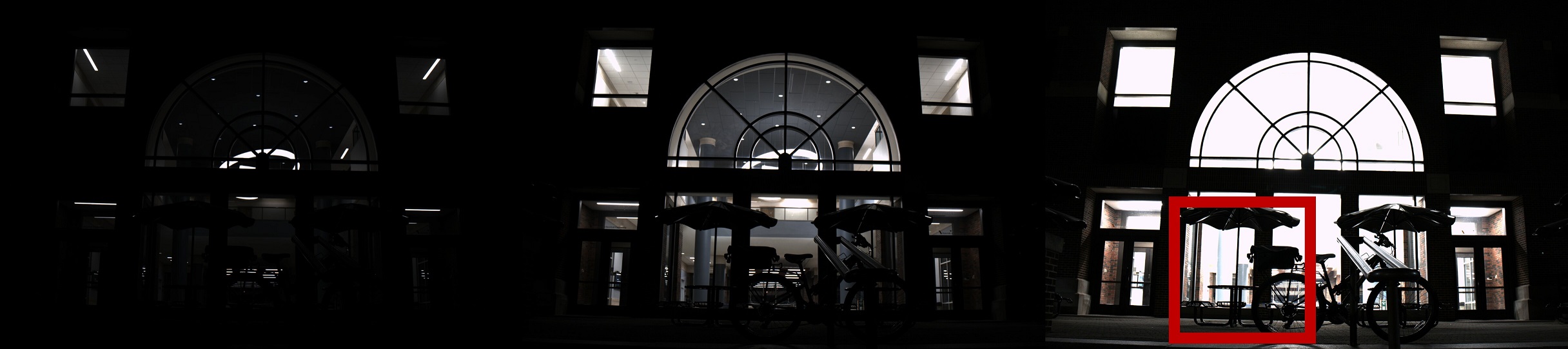} \\
  \vspace{-0.3ex}
    \subfloat[Input / \textbf{ISO 12800}]{%
    \includegraphics[width=0.165\linewidth]{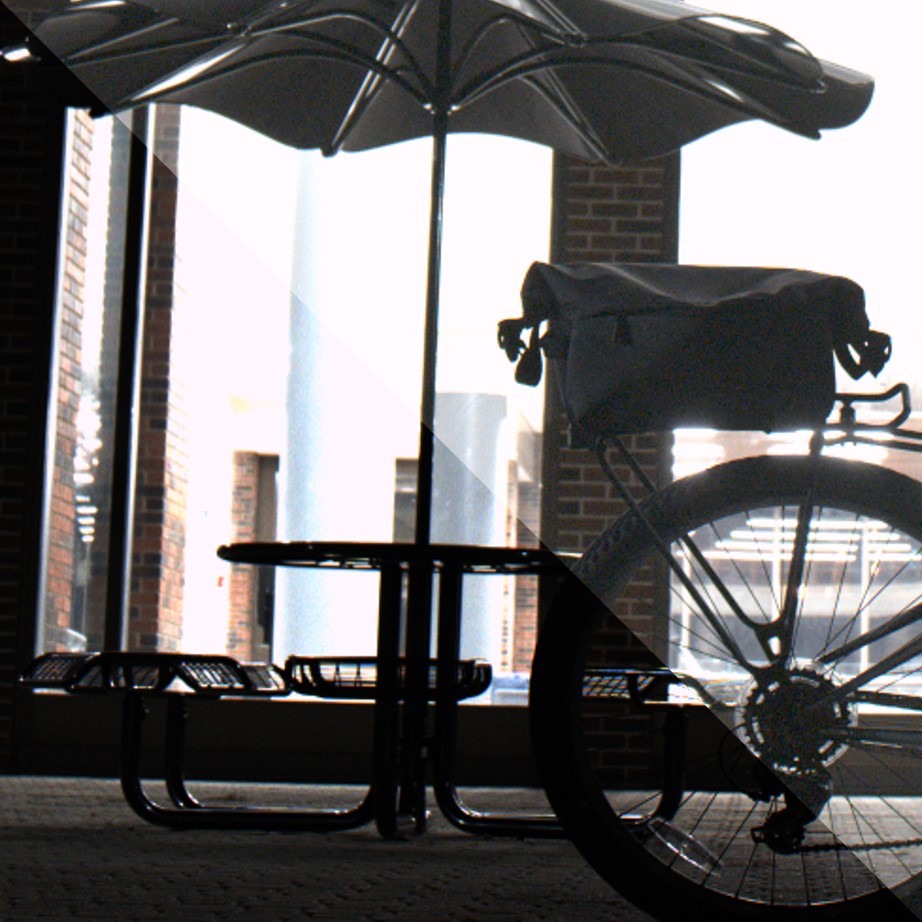}}
    \hfill
    \subfloat[Kalantari \cite{kalantari2017deep}]{%
    \includegraphics[width=0.165\linewidth]{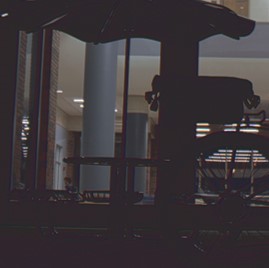}}
   \hfill
    \subfloat[Wu \cite{wu2018deep}]{%
    \includegraphics[width=0.165\linewidth]{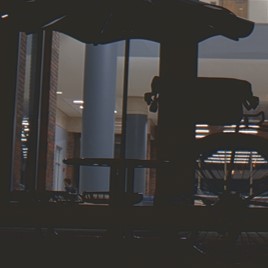}}
    \hfill
    \subfloat[AHDRNet \cite{yan2019attention}]{%
    \includegraphics[width=0.165\linewidth]{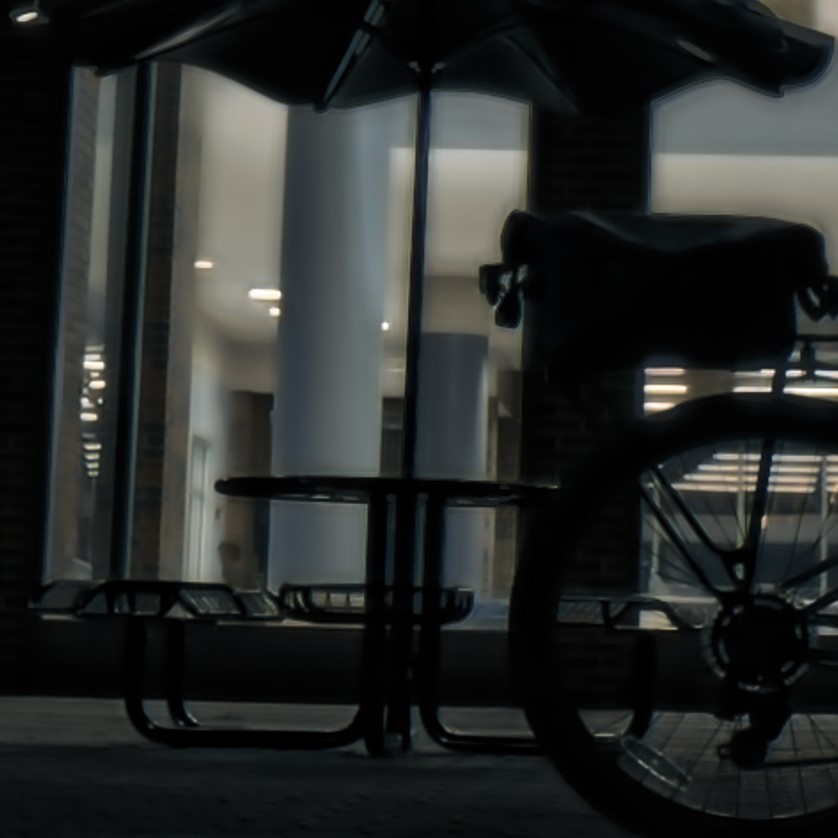}}
     \hfill
  \subfloat[NHDRRNet \cite{yan2020deep}]{%
    \includegraphics[width=0.165\linewidth]{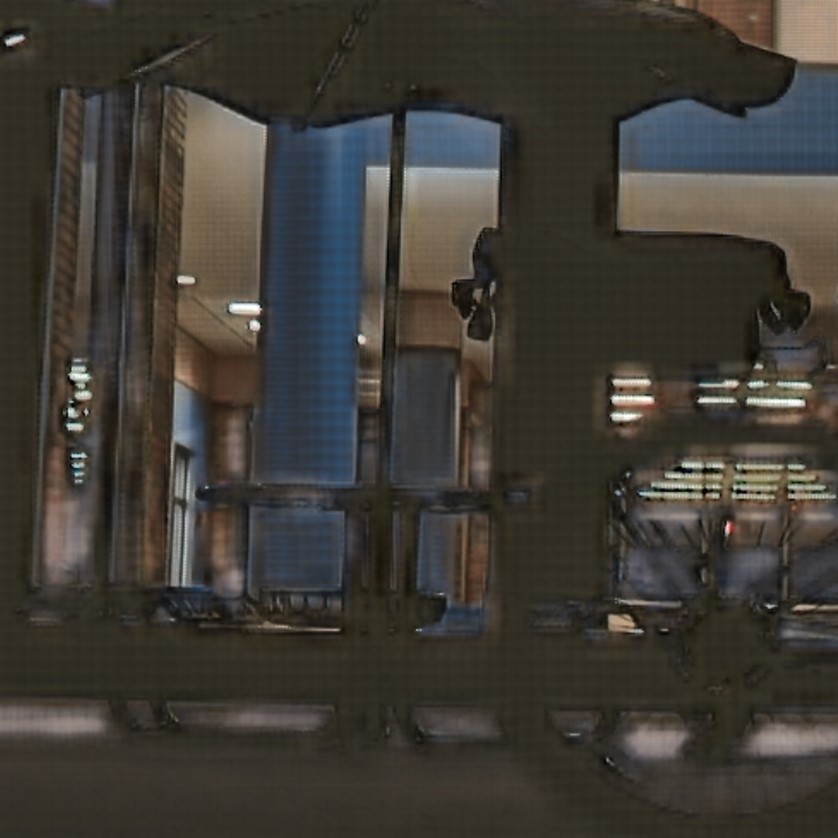}}
   \hfill
  \subfloat[\textbf{SV-HDR [Ours]}]{%
    \includegraphics[width=0.165\linewidth]{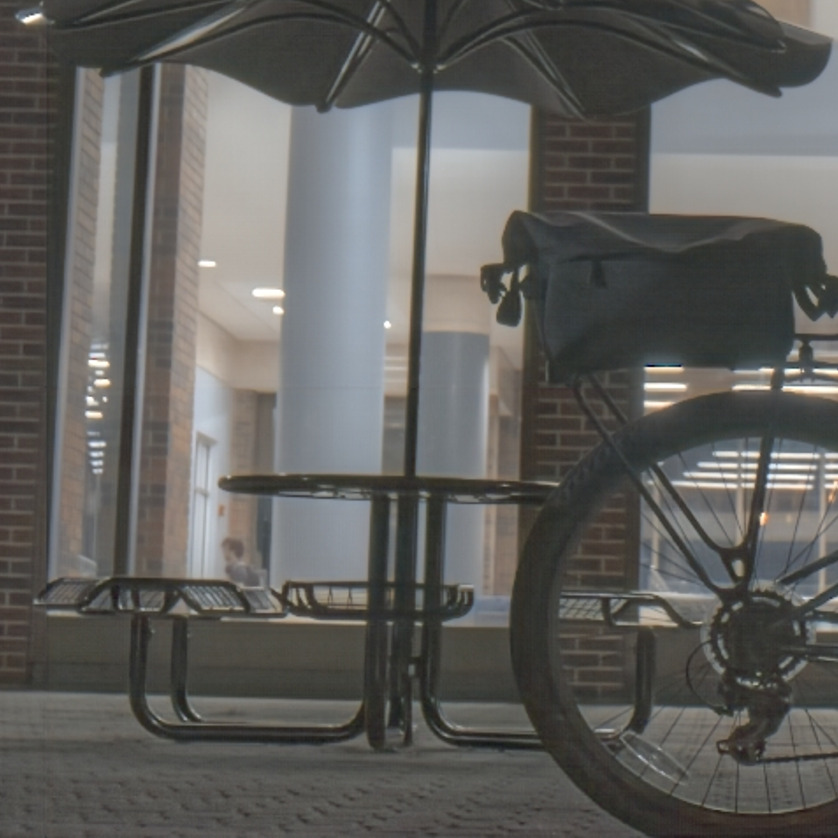}}\\
  \caption{Qualitative comparison using real data. [Top] Real images captured at different exposure times. [Bottom] Real images captured with different ISO. Gamma-corrected (top right) and raw (bottom left) images are shown in (a) and (f) to highlight the presence of spatially varying noise.}
  \label{fig: Figure18 Real Experiment}
\end{figure*}

\subsection{Ablation Study} \label{ablation}

\textbf{MSA Module}. We conducted an ablation study on types of multi-head self-attention (MSA) modules to build our SV-HDR network. Each MSA module is followed by an MLP. The candidates of MSA are: Window-based MSA (W-MSA) \cite{Wang_2022_CVPR}, Shifted Window-based MSA (SW-MSA) \cite{liang2021swinir}, and multi-Dconv head transposed attention (MDTA) \cite{zamir2022restormer}. We further experimented with different MSA choices for encoder and decoder. We test all configurations on the Kalantari test dataset \cite{kalantari2017deep} at a fixed $\tau$ of 4. Results are shown in Table \ref{tab: transformer ablation} (Top). We empirically found that SW-MSA performs the best among all candidates in both the encoder and decoder. We also found using channel attention as a feature extractor can lead to performance degrades.

\textbf{Feed-Forward Module}. We also ablate on types of feed-forward modules that follow MSA, while fixing SW-MSA as the spatial attention. The candidates of feed-forward are MLP, Locally-enhanced Feed-Forward Network (LeFF) \cite{Wang_2022_CVPR}, Depth-wise Locally-enhanced Feed-Forward Network (DwLeFF), and Gated-Dconv feed-forward network (GDFN) \cite{zamir2022restormer}. Results are shown in Table \ref{tab: transformer ablation} (Bottom). We found MLP outperforms all other feed-forward modules.

\textbf{Expo-Share Block}. We further compared the performance of SV-HDR with the deformable convolutions in the Expo-Share blocks replaced by standard convolutions. We plot the average test PSNR against illuminance for both models in \fref{fig: Figure15 dc ablation}. SV-HDR with deformable convolutions performs consistently better than with standard convolutions only.

\section{Conclusion}

HDR image fusion algorithms today typically lack the capability to handle noise. They tend to fail in low-light conditions, where signals will be contaminated heavily by the photon shot noise. In this paper, we provided evidence to show the existence of the problem. We found that the issue is caused by the co-presence of heavy noise and a wide dynamic range. A new HDR fusion and denoising network (SV-HDR) is presented as a solution to the problem. By introducing a customized multi-scale transformer and a new exposure-share block, we demonstrated the possibility of fusing noisy images with real cameras. 

\textbf{Acknowledgement}. This work is supported, in part, by the US National Science Foundation under the grants ECCS-2030570, IIS-2133032, DMS-2134209, a gift from Google, and a gift from Intel Labs. The authors thank Dr. Vladlen Koltun for his continuous support of this project and valuable advice.

\bibliographystyle{ieee_fullname}

\end{document}